\documentclass[11pt,a4paper]{article}
\usepackage[american]{babel}
\usepackage{amsmath,babel}
\usepackage{amssymb,latexsym}
\usepackage{amstext}
\usepackage{graphicx}
\usepackage{epsfig}
\usepackage[latin1]{inputenc}
\usepackage{amsfonts}
\usepackage{xcolor}

\newtheorem{theorem}{Theorem}

\newtheorem{definition}[theorem]{Definition}
\newtheorem{example}[theorem]{Example}

\newtheorem{lemma}[theorem]{Lemma}

\newtheorem{proposition}[theorem]{Proposition}

\newenvironment{proof}[1][Proof]{\noindent\textbf{#1.} }{$\Box$\\}

\usepackage[authoryear,round]{natbib}
\usepackage{longtable,booktabs,array}
\usepackage[top=0.5cm,bottom=0.5cm,right=2cm,left=2cm,headsep=0.4in,headheight=2.35cm,includeheadfoot]{geometry}
\newcommand{\HD}{\varphi^{hd}}
\newcommand{\HDi}{H_i}
\newcommand{\HDj}{H_j}
\newcommand{\HDip}{H_i^{\prime}}
\newcommand{\HDjp}{H_j^{\prime}}
\newcommand{\HDipp}{H_i^{\prime\prime}}

\newcommand{\HDkp}{H_k^{\prime}}
\newcommand{\HDkpp}{H_k^{\prime\prime}}
\newcommand{\HDl}{H_l}
\newcommand{\HDlp}{H_l^{\prime}}
\newcommand{\HDilp}{H^{\prime}_{i_{l}}}
\newcommand{\HDilpp}{H^{\prime\prime}_{i_{l}}}
\newcommand{\HDjhpp}{H^{\prime\prime}_{j_{h}}}
\newcommand{\HDjqpp}{H^{\prime\prime}_{q+1}}

\newcommand{\CDi}{C_i}
\newcommand{\CDj}{C_j}
\newcommand{\CDip}{C_i^{\prime}}
\newcommand{\CDjp}{C_j^{\prime}}
\newcommand{\CDipp}{C_i^{\prime\prime}}

\newcommand{\CDkp}{C_k^{\prime}}
\newcommand{\CDkpp}{C_k^{\prime\prime}}
\newcommand{\CDl}{C_l}
\newcommand{\CDlp}{C_l^{\prime}}

\newcommand{\CDilp}{C^{\prime}_{i_{l}}}

\newcommand{\CDjhpp}{C^{\prime\prime}_{j_{h}}}

\renewcommand{\baselinestretch}{1.5}

\begin{document}
	
\title{Evaluating the impact of items and cooperation in inventory models with exemptable ordering costs}
\author{M.G. Fiestras-Janeiro$^1$ fiestras@uvigo.es\\ 
	I. Garc\'{\i}a-Jurado$^{2,5}$ ignacio.garcia.jurado@udc.es \\
	A. Meca$^3$ ana.meca@umh.es\\
	M.A. Mosquera$^{4,5}$ mamrguez@uvigo.es}
\date{{\empty}}
\maketitle
\footnotetext[1]{Universidade de Vigo, Departamento de Estat\'{\i}stica e Investigaci\'on Operativa, 36310 Vigo, Spain.} 
\footnotetext[2]{Universidade da Coru\~{n}a, Departamento de Matem\'aticas, Grupo MODES,  15071 A Coru\~{n}a, Spain.}
\footnotetext[3]{Universidad Miguel Hernández de Elche, I.U. Centro de Investigación Operativa, 03202 Elche, Spain.} 
\footnotetext[4]{Universidade de Vigo, Departamento de Estat\'{\i}stica e Investigaci\'on Operativa, 32004 Ourense, Spain.} 
\footnotetext[5]{CITMAga, 15782 Santiago de Compostela, Spain.}

\begin{abstract}
In this paper we introduce and analyse, from a game theoretical perspective, several multi-agent or multi-item continuous review inventory models in which the buyers are exempted from ordering costs if the price of their orders is greater than or equal to a certain amount. For all models we obtain the optimal ordering policy. We first analyse a simple model with one firm and one item. Then, we study a model with one firm and several items, for which we design a procedure based on cooperative game theory to evaluate the impact of each item on the total cost. Then, we deal with a model with several firms and one item for each firm, for which we characterise a rule to allocate the total cost among the firms in a coalitionally stable way. Finally, we discuss a model with several firms and several items, for which we characterise a rule to allocate the total cost among the firms in a coalitionally stable way and to evaluate the impact of each item on the cost that would be payable to each firm when using the  allocation rule. All the concepts and results of this article are illustrated using data from a case study.
\end{abstract}
	

\noindent
{\bf Corresponding author:} I. Garc\'{\i}a-Jurado

\section{Introduction}
The inventory models described in this article are based on the activity of an electrical material distribution firm that contacted one of the authors for advice. This firm has warehouses in industrial estates in several cities where it stores and sells the electrical material requested by its customers, mainly professional electricians (this material is previously acquired by the firm from one or several suppliers). The firm competes in each city with other firms with similar characteristics. Its customers have, in general, considerable loyalty, although they expect that each time they make a purchase of regular items, such items will be immediately available. The warehouses are sufficiently spacious so that the firm has no capacity constraints. A common characteristic of most of the items it distributes is that they have a regular and predictable demand that can therefore be considered deterministic and linear in time. Thus, the firm faces in each of the cities and with each of the items an Economic Order Quantity (EOQ) inventory problem with no allowable shortages.

In inventory models there are typically two types of costs: ordering costs and holding costs. A specific feature of the situations that motivate this work is that their ordering costs are essentially shipping costs and that suppliers usually exempt the firm from such ordering costs when the orders it places are sufficiently large. Therefore, the problems that the firm addresses are of a new type that we call {\em EOQ problems with exemptable ordering costs}.  In the inventory literature there are a number of papers that consider reductions in the costs applied by suppliers that, in some ways, have some points of similarity with our problem. Below is a brief survey of such papers.

\cite{Baumol1970} considered a model where all ordering costs are shipping costs; the emphasis in this article was on the choice of the most convenient way of transport. \cite{Langley1981} introduced a shipping cost as an additional element in the EOQ model.	\cite{Aucamp1982}  made the shipping cost dependent on the number of trucks needed for transportation,  but without considering discounts. \cite{Lee1986} introduced  discounts in the ordering cost associated to freight transportation. In particular this paper considered that the ordering cost is divided into two parts: a fixed part plus a variable part representing shipping costs, which is a decreasing function of the size of the order. An algorithm to obtain the optimal order quantity is provided. \cite{Lee1989} considered  the dynamic  lot size problem where the shipping costs are proportional to the number of containers used. \citet{Hwang1990} extended the model in \cite{Lee1986} by adding also  discounts on purchase prices. 
\cite{Tersine1991} analysed EOQ situations with shipping discounts: all-units or  incremental quantity discounts, and all-weight or incremental freight discounts. \cite{Aucamp1984} analysed a model where the cost per order has three components: a fixed part, a transportation part which depends on the quantity ordered, and the purchasing charge. \cite{Shinn1996} assumed the Lee's  ordering cost structure and also allowed delay in payments. \cite{Burwell1997} incorporated quantity and shipping discounts when demand depends on the purchasing price. \cite{Toptal2009} extended the model in 	\cite{Aucamp1982} using a general replenishment cost structure which includes stepwise shipping costs and all-units quantity discounts. \citet{Frenk2014} discussed the EOQ model for one firm considering that the ordering cost has a fixed part plus a variable part depending on the ordered quantity. \cite{Bigham1986} and \cite{Gupta1994} also considered inventory models in which ordering costs depend on order size but, contrary to our work and all those just reviewed, such dependence is not decreasing but increasing. \cite{Pereira2015} offered a survey on EOQ models with incremental and all-units discounts.

In the last two decades, EOQ models have been extended to the multi-agent case. \citet{Meca2004}  is among the first works to consider cooperation in EOQ models when several firms place joint orders. Since then, many articles have analysed this type of models halfway between inventory theory and cooperative game theory; \cite{Fiestras2011} provided a survey of such literature. Recently, \citet{Li2021} extended the Meca et al.'s  model to situations where the supplier offers firms a price discount on purchases above a certain order quantity. Li's approach is the closest to that of our article, because it combines cooperation between several firms and discounts that depend on the size of the order. However, Li's discounts are applied to the unit acquisition costs of the item, while ours are applied to the fixed cost of the order. Moreover, Li's model refers to a single item, whereas we consider  multi-item models in which cooperation can occur not only between firms but also between items. Even though we propose a cooperation among items, in our model the ordering cost is the same for all items in contrast to the case of the joint replenishment model where the ordering cost depends on the item type. Besides,  in the literature on joint replenishment,  there are models that consider certain types of discounts: depending on the number of packages \citep{Goyal1973}, allowing quantity discounts \citep{Benton1991},  making the ordering cost dependent on the total purchase value   \citep{Xu2000}, etc.   \cite{Khouja2008} offered a wide survey of the models and algorithms applied to this problem. \cite{Peng2022} presented an updated review, including papers that explored the gain splitting under cooperation. Nevertheless, the case where the ordering cost  vanishes has not yet been  studied as far as we know.

The structure of the paper is the following. In Section 2 we formulate a basic EOQ problem with exemptable ordering costs, for which we obtain the optimal ordering policy. In Section 3 we study a model with one firm and several items, for which we obtain an optimal ordering policy and a procedure based on cooperative game theory to evaluate the impact of each item on the total optimal cost; our procedure makes use of the Shapley value. In Section 4 we deal with a model with several firms and one item for each firm, for which we obtain an optimal ordering policy and a rule to allocate the total cost among the firms in a coalitionally stable way; we also provide an axiomatic characterization of our rule. In Section 5 we discuss
 a model with several firms and several items, for which we characterise a rule to allocate
 the total cost among the firms in a coalitionally stable way and  to evaluate
 the impact of each item on the cost that would be payable to each firm when using the allocation rule. The concepts and results of this article are illustrated with data from a case study that we develop in Sections 3, 4 and 5. Finally, Section 6 draws conclusions and identifies further research for scholars in the field.

\section{A basic EOQ problem with exemptable ordering costs}
In a basic EOQ problem with exemptable ordering costs a single firm has to satisfy the demand for a product that it sells to its customers. To do so, it buys this product from a supplier and stores its stock in a warehouse of unlimited capacity. It can place orders with the supplier at any time; the supplier waives the ordering costs if the order is sufficiently large. The firm does not allow shortages. Demand is assumed to be deterministic and linear. The lead time (the time between placement of an order and delivery of that order), is assumed to be deterministic and constant and, without loss of generality, equal to zero. The parameters that characterise such a problem are as follows:

\begin{itemize}
	\item $d>0$ is the demand of the product per unit of time.
	\item $h>0$ is the cost of holding one unit of product for one unit of time.
	\item $a>0$ is the ordering cost.
	\item $A> 0$ is the order size above which the supplier exempts the firm from the ordering cost.
\end{itemize}
The decision variable in this problem is $Q$, the size of the order to be placed by the firm as soon as the inventory level of the product becomes zero. We will now calculate the order size that minimises the average cost per unit of time. We assume that the acquisition cost of each unit of the product does not depend on the size of the order. Therefore, since all demand must be satisfied, the acquisition cost of the product is not relevant to the optimisation problem. The average cost per cycle (the period between two orders) is given by
\begin{equation}
\label{costpercycle}
CPC(Q)=\left\{
\begin{array}{ll}
a+h\frac{Q}{2}\frac{Q}{d}&\mbox{if $Q<A$,}\\\\
h\frac{Q}{2}\frac{Q}{d}&\mbox{if $Q\geq A$.}\\
\end{array}
\right.
\end{equation}
Now, since the cycle length is $\frac{Q}{d}$, the average cost per time unit is given by
\begin{equation}
\label{costpertimeunit}
CPT(Q)=\frac{CPC(Q)}{\frac{Q}{d}}=\left\{
\begin{array}{ll}
\frac{ad}{Q}+h\frac{Q}{2}&\mbox{if $Q<A$,}\\\\
h\frac{Q}{2}&\mbox{if $Q\geq A$.}\\
\end{array}
\right.
\end{equation}
In view of (\ref{costpertimeunit}), it is easy to check after some algebra that the unique minimum of $CPT$ is $\bar{Q}$ given by
\begin{equation}
\label{minimumCPT}
\bar{Q}=\left\{
\begin{array}{ll}
\sqrt{\frac{2ad}{h}}&\mbox{if $2\sqrt{\frac{2ad}{h}}<A$,}\\\\
A&\mbox{if $2\sqrt{\frac{2ad}{h}}\geq A$.}\\
\end{array}
\right.
\end{equation}
Furthermore the minimum average cost per time unit is given by
\begin{equation}
\label{minimumvalueofCPT}
CPT(\bar{Q})=\left\{
\begin{array}{ll}
\sqrt{2adh}&\mbox{if $2\sqrt{\frac{2ad}{h}}<A$,}\\\\
\frac{hA}{2}&\mbox{if $2\sqrt{\frac{2ad}{h}}\geq A$.}\\
\end{array}
\right.
\end{equation}
Equivalently, $CPT(\bar{Q})=\min\{\sqrt{2adh},\frac{hA}{2}\}$.
\begin{example} Consider the basic EOQ problem with exemptable ordering costs characterised by $d=15$, $h=8$, $a=10$ and $A=10$. The corresponding function $CPT$, according to (\ref{costpertimeunit}), is given by
\begin{equation}
\label{costpertimeunit_ex1}
CPT(Q)=\left\{
\begin{array}{ll}
\frac{150}{Q}+4Q&\mbox{if $Q<10$,}\\\\
4Q&\mbox{if $Q\geq 10$.}\\
\end{array}
\right.
\end{equation}
Figure \ref{one} displays the function $CPT$. It clearly shows that $\bar{Q}=10$ and $CPT(\bar{Q})=40$. If we calculate the minimum using  (\ref{minimumCPT}) we obviously arrive at the same result. Note that $CPT$ is neither continuous nor convex.
\begin{figure}
\label{one}
\centerline{\includegraphics[scale=0.5]{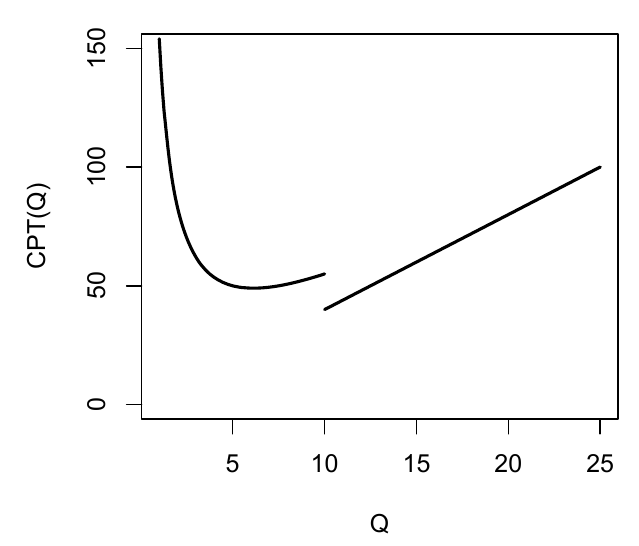}}
\caption{The function $CPT$ for $d=15$, $h=8$, $a=10$, $A=10$.}
\end{figure}
\end{example}
\section{A problem with multiple items}
\label{sectionthree}
In the previous section we looked at the simplest case where the firm considers its optimal order size problem on an item-by-item basis. However, a distribution firm such as the one in question is likely to deal with several items for each of its suppliers. In such a case the exemption from ordering costs when the order is sufficiently large will depend on the price of the order exceeding a certain quantity.  This gives rise to the {\em multi-item EOQ problem with exemptable ordering costs} discussed below. The parameters that characterise such a problem are as follows:

\begin{itemize}
	\item $N$ is the finite set of items.
	\item $d_i>0$ is the demand of item $i\in N$ per unit of time.
	\item $h_i>0$ is the cost of holding one unit of item $i\in N$ for one unit of time.
	\item $c_i>0$ is the acquisition cost of one unit of item $i\in N$.
	\item $a>0$ is the ordering cost.
	\item $B> 0$ is the order price above which the supplier exempts the firm from the ordering cost.
\end{itemize}
We assume that the firm has decided to place joint orders of items in $N$. This means that, if we denote by $Q_i$ the order size of item $i$, it holds that
\begin{equation}
\label{cycles_equality}
\frac{Q_i}{d_i}=\frac{Q_j}{d_j}, \mbox{ for all $i,j\in N$}.
\end{equation}
The average cost per cycle for a given collection $S\subseteq N$ is now given by
\begin{equation}
\label{costpercycle+}
CPC(\{Q_j\}_{j\in S})=\left\{
\begin{array}{ll}
a+\sum_{j\in S}h_j\frac{Q_j}{2}\frac{Q_j}{d_j}&\mbox{if $\sum_{j\in S}c_jQ_j<B$,}\\\\
\sum_{j\in S}h_j\frac{Q_j}{2}\frac{Q_j}{d_j}&\mbox{if $\sum_{j\in S}c_jQ_j\geq B$.}\\
\end{array}
\right.
\end{equation}
Now, since the cycle length is $\frac{Q_i}{d_i}$ for any $i\in S$ (see (\ref{cycles_equality})), the average cost per time unit is given by
\begin{equation}
\label{costpertimeunit+}
CPT(\{Q_j\}_{j\in S})=\left\{
\begin{array}{ll}
\frac{ad_i}{Q_i}+\sum_{j\in S}h_j\frac{Q_j}{2}&\mbox{if $\sum_{j\in S}c_jQ_j<B$,}\\\\
\sum_{j\in S}h_j\frac{Q_j}{2}&\mbox{if $\sum_{j\in S}c_jQ_j\geq B$.}\\
\end{array}
\right.
\end{equation}
Expression (\ref{cycles_equality}) implies that, if we fix $i\in S$, then $Q_j=d_j\frac{Q_i}{d_i}$ for all $j\in S$. Hence, $CPT$ can be seen as a function of $Q_i$ by writing:
\begin{equation}
\label{costpertimeunit++}
CPT(Q_i)=\left\{
\begin{array}{ll}
\frac{ad_i}{Q_i}+\frac{Q_i}{2d_i}\sum_{j\in S}h_jd_j&\mbox{if $\frac{Q_i}{d_i}\sum_{j\in S}c_jd_j<B$,}\\\\
\frac{Q_i}{2d_i}\sum_{j\in S}h_jd_j&\mbox{if $\frac{Q_i}{d_i}\sum_{j\in S}c_jd_j\geq B$.}\\
\end{array}
\right.
\end{equation}
In view of (\ref{costpertimeunit++}), it is easy to check after some algebra that the unique minimum of $CPT$ is $\hat{Q_i}$ given by
\begin{equation}
\label{minimumCPT+}
\hat{Q_i}=\left\{
\begin{array}{ll}
\sqrt{\frac{2ad_i^2}{\sum_{j\in S}h_jd_j}}&\mbox{if $2\sqrt{\frac{2a}{\sum_{j\in S}h_jd_j}}<\frac{B}{\sum_{j\in S}c_jd_j}$,}\\\\
\frac{B d_i}{\sum_{j\in S}c_jd_j}&\mbox{if $2\sqrt{\frac{2a}{\sum_{j\in S}h_jd_j}}\geq\frac{B}{\sum_{j\in S}c_jd_j}$,}\\
\end{array}
\right.
\end{equation}
for every $i\in S$. Furthermore the minimum average cost per time unit is given by
\begin{equation}
\label{minimumvalueofCPT+}
CPT(\hat{Q}_i)=\left\{
\begin{array}{ll}
\sqrt{2a\sum_{j\in S}h_jd_j}&\mbox{if $2\sqrt{\frac{2a}{\sum_{j\in S}h_jd_j}}<\frac{B}{\sum_{j\in S}c_jd_j}$,}\\\\
\frac{B}{2}\frac{\sum_{j\in S}h_jd_j}{\sum_{j\in S}c_jd_j}&\mbox{if $2\sqrt{\frac{2a}{\sum_{j\in S}h_jd_j}}\geq\frac{B}{\sum_{j\in S}c_jd_j}$.}\\
\end{array}
\right.
\end{equation}
or, equivalently, by
\begin{equation}
	\label{minimumvalueofCPT+-}
	CPT(\hat{Q}_i)=\min\left\{\sqrt{2a\sum_{j\in S}h_jd_j},\frac{B}{2}\frac{\sum_{j\in S}h_jd_j}{\sum_{j\in S}c_jd_j}\right\}
\end{equation}
Notice that $CPT(\hat{Q}_i)$ is the total cost per unit of time incurred by items of $S$ and does not really depend on a particular $i$. Therefore, we can denote this quantity interchangeably by $CPT(\hat{Q}_i)$ or by $CPT(\{\hat{Q}_j\}_{j\in S})$. Now, for every multi-item EOQ problem with exemptable ordering costs 
$$P=(N, \{d_i\}_{i\in N},\{h_i\}_{i\in N},\{c_i\}_{i\in N},a,B )$$
we can define its associated cost game $(N,c^P)$ given by
\begin{equation}\label{eq:game_item}
	c^P(S)=CPT(\{\hat{Q}_j\}_{j\in S})
\end{equation}
for all $S\subseteq N$. Observe that (\ref{minimumvalueofCPT+-}) implies that, for all $S\subseteq N$,

\begin{equation}\label{eq:game_item2}
	c^P(S)= CPT(\{\hat{Q}_j\}_{j\in S})
	= \left(\sum_{j\in S}h_jd_j\right) \min\left\{\dfrac{B}{2\sum_{j\in S}c_jd_j},\sqrt{\dfrac{2a}{\sum_{j\in S}h_jd_j}}\right\}.
\end{equation}
The cost game $(N,c^P)$ is said to be \textit{strictly subadditive} if, for each pair of non-empty coalitions $S,T\subseteq N$ such that $S\cap T =\emptyset$, it holds that $c^P(S\cup T) < c^P(S)+ c^P(T)$. Note that strict subadditivity implies that it is reasonable for the grand coalition $N$ to form (i.e., it is reasonable for the firm to place joint orders for items of $N$), since this minimises the total cost.

\begin{proposition}
	Let $P = (N,\{d_i\}_{i\in N},\{h_i\}_{i\in N},\{c_i\}_{i\in N},a,B)$ be a multi-firm EOQ problem with exemptable ordering costs and let $(N, c^P)$ be its associated cost game. Then, $(N, c^P)$ is strictly subadditive.
\end{proposition}

\begin{proof}
	Take a pair of non-empty coalitions $S,T\subseteq N$ such that $S\cap T =\emptyset$ and see that $c^P(S\cup T) < c^P(S)+ c^P(T)$. Indeed, using (\ref{eq:game_item2})
	\begin{eqnarray*}
		c^P(S\cup T)&=& \left(\sum_{j\in S\cup T}h_jd_j\right) \min\left\{\dfrac{B}{2\sum_{j\in S\cup T}c_jd_j},\sqrt{\dfrac{2a}{\sum_{j\in S\cup T}h_jd_j}}\right\}\\
		&=& \left(\sum_{j\in S}h_jd_j\right) \min\left\{\dfrac{B}{2\sum_{j\in S\cup T}c_jd_j},\sqrt{\dfrac{2a}{\sum_{j\in S\cup T}h_jd_j}}\right\}\\
		& +& \left(\sum_{j\in T}h_jd_j\right) \min\left\{\dfrac{B}{2\sum_{j\in S\cup T}c_jd_j},\sqrt{\dfrac{2a}{\sum_{j\in S\cup T}h_jd_j}}\right\}\\
		&<& \left(\sum_{j\in S}h_jd_j\right) \min\left\{\dfrac{B}{2\sum_{j\in S}c_jd_j},\sqrt{\dfrac{2a}{\sum_{j\in S}h_jd_j}}\right\}\\
		& +& \left(\sum_{j\in T}h_jd_j\right) \min\left\{\dfrac{B}{2\sum_{j\in T}c_jd_j},\sqrt{\dfrac{2a}{\sum_{j\in T}h_jd_j}}\right\}\\
		&=& c^P(S) +c^P(T),
	\end{eqnarray*}
	where the inequality follows from $\sum_{j\in S \cup T}c_jd_j > \sum_{j\in S}c_jd_j>0$, $\sum_{j\in S \cup T}c_jd_j > \sum_{j\in T}c_jd_j>0$, $\sum_{j\in S \cup T}h_jd_j > \sum_{j\in S}h_jd_j>0$, and $\sum_{j\in S \cup T}h_jd_j > \sum_{j\in T}h_jd_j>0$.
\end{proof}

 For various organisational or management reasons, it may be in the firm's interest to spread the cost per unit of time fairly among the various items; this gives, for example, an idea of the impact that each of these items has in the firm's inventory costs\footnote{Throughout this article, when we write {\em inventory costs}, we refer to the sum of the ordering costs plus the holding costs resulting from the inventory policy used.}. One tool borrowed from cooperative game theory that we use to spread such cost in a fair way is the Shapley value.  The Shapley value was introduced in \cite{Shapley1953} and, since then, has been widely used to propose fair distributions of the benefits of cooperation among various agents. When we speak of {\em fair} distributions, we mean that they are the result of adopting a collection of axioms that mathematically express ideas having to do with justice. We can find numerous axiomatic characterizations of the Shapley value in the literature (see, e.g., \cite{Gonzalez-Diaz2010} for an introduction to cooperative game theory and the Shapley value), which show its worth as a tool for producing fair distributions from different points of view.
 
 Specifically, we define a rule that maps the vector $SH(N,c^P)\in\mathbb{R}^N$ to each problem $P$, where $SH$ denotes the Shapley value. Thus, for every item $i\in N$, the cost allocated to $i$  is given by
$$
SH_i(N,c^P)=\sum_{S\subseteq N\setminus \{i\}}\frac{|S|!(|N|-|S|-1)!}{|N|!}\left(c^P(S\cup\{i\})-c^P(S)\right).
$$
	
Below is an example showing how to determine the optimal order size in the multi-item case and how to use the Shapley value to extract information about the impact of the various items on the inventory cost.

\begin{example}
	\label{ex:2}
We consider a firm that sells to its customers one hundred items that it previously buys from a unique supplier. Because of the characteristics of the firm, its supplier and its customers, the corresponding inventory problem fits into a multi-item EOQ model with exemptable ordering costs. The supplier is willing to meet any order placed by the firm at an order cost of EUR $2\,000$. The firm is exempted from this charge for all orders greater than or equal to EUR $200\,000$. Table \ref{t1} shows for each of the items their corresponding parameters, i.e. their monthly demands, their holding costs per unit and month, and their acquisition costs per unit. 

\renewcommand{\baselinestretch}{1}

\begin{table}
	\fontsize{8}{10}\selectfont
	\begin{tabular}[t]{rrrr}
		\toprule
		Item & $d$ & $h$ & $c$\\
		\midrule
		1 & 419 & 0.45 & 4.03\\
		2 & 467 & 0.46 & 4.13\\
		3 & 183 & 0.32 & 94.01\\
		4 & 18 & 0.23 & 68.20\\
		5 & 199 & 0.12 & 64.41\\
		\addlinespace
		6 & 430 & 0.47 & 1.18\\
		7 & 310 & 0.19 & 5.53\\
		8 & 122 & 0.08 & 58.53\\
		9 & 303 & 0.48 & 8.84\\
		10 & 233 & 0.37 & 58.61\\
		\addlinespace
		11 & 248 & 0.11 & 74.10\\
		12 & 18 & 0.30 & 62.65\\
		13 & 378 & 0.48 & 1.06\\
		14 & 157 & 0.31 & 60.81\\
		15 & 94 & 0.23 & 83.72\\
		\addlinespace
		16 & 95 & 0.34 & 52.38\\
		17 & 260 & 0.19 & 85.04\\
		18 & 201 & 0.19 & 67.59\\
		19 & 95 & 0.15 & 70.45\\
		20 & 445 & 0.22 & 1.65\\
		\addlinespace
		21 & 352 & 0.49 & 2.48\\
		22 & 141 & 0.12 & 91.05\\
		23 & 481 & 0.09 & 7.93\\
		24 & 359 & 0.11 & 7.62\\
		25 & 332 & 0.36 & 9.75\\
		\bottomrule
	\end{tabular}
	\hfill
	\begin{tabular}[t]{rrrr}
		\toprule
		Item & $d$ & $h$ & $c$\\
		\midrule
		26 & 30 & 0.33 & 95.94\\
		27 & 11 & 0.45 & 64.13\\
		28 & 430 & 0.35 & 5.20\\
		29 & 141 & 0.38 & 98.06\\
		30 & 258 & 0.28 & 86.42\\
		\addlinespace
		31 & 215 & 0.35 & 84.32\\
		32 & 424 & 0.42 & 1.67\\
		33 & 82 & 0.40 & 52.64\\
		34 & 85 & 0.49 & 69.76\\
		35 & 47 & 0.25 & 73.89\\
		\addlinespace
		36 & 363 & 0.19 & 6.84\\
		37 & 377 & 0.23 & 7.83\\
		38 & 336 & 0.05 & 2.23\\
		39 & 147 & 0.13 & 78.01\\
		40 & 36 & 0.43 & 84.91\\
		\addlinespace
		41 & 494 & 0.15 & 4.57\\
		42 & 113 & 0.16 & 95.78\\
		43 & 267 & 0.08 & 80.92\\
		44 & 397 & 0.16 & 3.02\\
		45 & 429 & 0.38 & 1.52\\
		\addlinespace
		46 & 334 & 0.43 & 4.56\\
		47 & 27 & 0.27 & 71.42\\
		48 & 415 & 0.22 & 1.58\\
		49 & 448 & 0.16 & 3.03\\
		50 & 313 & 0.10 & 1.49\\
		\bottomrule
	\end{tabular}
	\hfill
	\begin{tabular}[t]{rrrr}
		\toprule
		Item & $d$ & $h$ & $c$\\
		\midrule
		51 & 139 & 0.23 & 77.10\\
		52 & 313 & 0.31 & 7.03\\
		53 & 415 & 0.15 & 3.68\\
		54 & 228 & 0.25 & 52.92\\
		55 & 170 & 0.15 & 63.04\\
		\addlinespace
		56 & 221 & 0.28 & 69.86\\
		57 & 294 & 0.21 & 59.89\\
		58 & 481 & 0.34 & 1.91\\
		59 & 73 & 0.22 & 91.60\\
		60 & 76 & 0.21 & 57.64\\
		\addlinespace
		61 & 80 & 0.29 & 90.17\\
		62 & 67 & 0.38 & 77.34\\
		63 & 145 & 0.15 & 83.12\\
		64 & 214 & 0.24 & 58.58\\
		65 & 357 & 0.17 & 1.65\\
		\addlinespace
		66 & 351 & 0.33 & 8.92\\
		67 & 157 & 0.13 & 81.65\\
		68 & 298 & 0.44 & 65.59\\
		69 & 281 & 0.39 & 86.23\\
		70 & 467 & 0.35 & 7.79\\
		\addlinespace
		71 & 45 & 0.33 & 69.95\\
		72 & 435 & 0.22 & 8.35\\
		73 & 94 & 0.29 & 98.47\\
		74 & 320 & 0.44 & 9.84\\
		75 & 227 & 0.31 & 98.37\\
		\bottomrule
	\end{tabular}
	\hfill
	\begin{tabular}[t]{rrrr}
		\toprule
		Item & $d$ & $h$ & $c$\\
		\midrule
		76 & 20 & 0.43 & 86.34\\
		77 & 120 & 0.19 & 62.86\\
		78 & 98 & 0.37 & 61.09\\
		79 & 266 & 0.17 & 79.65\\
		80 & 239 & 0.32 & 63.38\\
		\addlinespace
		81 & 460 & 0.27 & 1.93\\
		82 & 90 & 0.17 & 76.55\\
		83 & 346 & 0.30 & 1.89\\
		84 & 43 & 0.46 & 89.26\\
		85 & 163 & 0.46 & 58.40\\
		\addlinespace
		86 & 244 & 0.17 & 70.22\\
		87 & 213 & 0.19 & 73.58\\
		88 & 378 & 0.49 & 8.19\\
		89 & 310 & 0.33 & 8.06\\
		90 & 38 & 0.47 & 93.41\\
		\addlinespace
		91 & 8 & 0.26 & 96.29\\
		92 & 17 & 0.23 & 94.10\\
		93 & 73 & 0.35 & 83.71\\
		94 & 387 & 0.12 & 1.08\\
		95 & 247 & 0.31 & 97.51\\
		\addlinespace
		96 & 413 & 0.16 & 8.01\\
		97 & 312 & 0.48 & 7.56\\
		98 & 282 & 0.32 & 75.82\\
		99 & 93 & 0.28 & 78.83\\
		100 & 420 & 0.23 & 6.67\\
		\bottomrule
	\end{tabular}
	
	\caption{\label{t1} Parameters for Example \ref{ex:2}.}
\end{table}

\renewcommand{\baselinestretch}{1.5}

First we obtain the optimal order plan for this firm taking into account that it will place joint orders for all items.  Using the expression (\ref{minimumCPT+}) we compute the optimal order size for each item, which can be seen in Table \ref{t2}; the resulting optimal cycle length is $0.2787$ months and therefore the optimal number of orders placed per month is $3.5868$. Table \ref{t2} also shows the parameters associated with all items, as well as the Shapley value\footnote{In this example, the Shapley value is approximated using the sampling methodology of \cite{Castro2009}.} of the corresponding game given by the expression (\ref{eq:game_item}). Note that the items are arranged in this table in increasing order of their Shapley values. The allocation of the Shapley value for each item gives a measure of the importance of each item's contribution to the joint inventory cost. This measure can help the inventory manager to make the right decisions in some circumstances. For example, if the firm is considering to stop distributing some items, their contributions to the joint inventory cost (as measured by the Shapley value) can be taken into account in making such a decision. Note that the Shapley value may propose negative allocations to some items. This makes sense because, indeed, the inclusion of some items in the joint order may reduce total inventory costs, given the special characteristics of the cost function of the problem we are dealing with. Finally, we can comment that, in view of the results presented in Table \ref{t2}, the items that contribute most to the joint inventory cost tend to have high demands and holding costs, but low acquisition costs.

\renewcommand{\baselinestretch}{1}

\begin{table}
	\fontsize{8}{10}\selectfont
	\begin{tabular}[t]{rrrrrrr}
		\toprule
	 & $d$ & $h$ & $c$ & $\hat{Q}_i$ & \multicolumn{1}{p{8ex}}{\raggedleft  Shapley value} & \multicolumn{1}{p{10ex}}{\raggedleft  hd-prop. value}\\
		\midrule
		43 & 267 & 0.08 & 80.92 & 74.44 & -76.38 & 2.98\\
		17 & 260 & 0.19 & 85.04 & 72.49 & -63.35 & 6.89\\
		79 & 266 & 0.17 & 79.65 & 74.16 & -61.83 & 6.30\\
		11 & 248 & 0.11 & 74.10 & 69.14 & -60.14 & 3.80\\
		95 & 247 & 0.31 & 97.51 & 68.86 & -56.99 & 10.67\\
		\addlinespace
		75 & 227 & 0.31 & 98.37 & 63.29 & -52.98 & 9.81\\
		30 & 258 & 0.28 & 86.42 & 71.93 & -51.73 & 10.07\\
		86 & 244 & 0.17 & 70.22 & 68.03 & -46.94 & 5.78\\
		22 & 141 & 0.12 & 91.05 & 39.31 & -43.08 & 2.36\\
		87 & 213 & 0.19 & 73.58 & 59.38 & -41.17 & 5.64\\
		\addlinespace
		67 & 157 & 0.13 & 81.65 & 43.77 & -40.86 & 2.85\\
		69 & 281 & 0.39 & 86.23 & 78.34 & -40.13 & 15.28\\
		5 & 199 & 0.12 & 64.41 & 55.48 & -38.71 & 3.33\\
		98 & 282 & 0.32 & 75.82 & 78.62 & -38.30 & 12.58\\
		3 & 183 & 0.32 & 94.01 & 51.02 & -37.55 & 8.16\\
		\addlinespace
		57 & 294 & 0.21 & 59.89 & 81.97 & -37.50 & 8.61\\
		63 & 145 & 0.15 & 83.12 & 40.43 & -36.79 & 3.03\\
		39 & 147 & 0.13 & 78.01 & 40.98 & -35.91 & 2.66\\
		42 & 113 & 0.16 & 95.78 & 31.50 & -33.78 & 2.52\\
		18 & 201 & 0.19 & 67.59 & 56.04 & -33.52 & 5.32\\
		\addlinespace
		31 & 215 & 0.35 & 84.32 & 59.94 & -32.18 & 10.49\\
		55 & 170 & 0.15 & 63.04 & 47.40 & -28.63 & 3.55\\
		56 & 221 & 0.28 & 69.86 & 61.61 & -27.76 & 8.63\\
		29 & 141 & 0.38 & 98.06 & 39.31 & -25.45 & 7.47\\
		51 & 139 & 0.23 & 77.10 & 38.75 & -24.57 & 4.46\\
		\addlinespace
		8 & 122 & 0.08 & 58.53 & 34.01 & -23.18 & 1.36\\
		73 & 94 & 0.29 & 98.47 & 26.21 & -20.97 & 3.80\\
		64 & 214 & 0.24 & 58.58 & 59.66 & -20.87 & 7.16\\
		15 & 94 & 0.23 & 83.72 & 26.21 & -18.37 & 3.01\\
		82 & 90 & 0.17 & 76.55 & 25.09 & -18.21 & 2.13\\
		\addlinespace
		80 & 239 & 0.32 & 63.38 & 66.63 & -18.08 & 10.66\\
		19 & 95 & 0.15 & 70.45 & 26.49 & -18.07 & 1.99\\
		59 & 73 & 0.22 & 91.60 & 20.35 & -16.74 & 2.24\\
		77 & 120 & 0.19 & 62.86 & 33.46 & -16.11 & 3.18\\
		54 & 228 & 0.25 & 52.92 & 63.57 & -15.40 & 7.95\\
		\addlinespace
		61 & 80 & 0.29 & 90.17 & 22.30 & -14.48 & 3.23\\
		99 & 93 & 0.28 & 78.83 & 25.93 & -13.21 & 3.63\\
		14 & 157 & 0.31 & 60.81 & 43.77 & -9.26 & 6.78\\
		93 & 73 & 0.35 & 83.71 & 20.35 & -8.34 & 3.56\\
		68 & 298 & 0.44 & 65.59 & 83.08 & -7.79 & 18.28\\
		\addlinespace
		60 & 76 & 0.21 & 57.64 & 21.19 & -7.01 & 2.22\\
		10 & 233 & 0.37 & 58.61 & 64.96 & -5.87 & 12.02\\
		35 & 47 & 0.25 & 73.89 & 13.10 & -5.83 & 1.64\\
		26 & 30 & 0.33 & 95.94 & 8.36 & -4.54 & 1.38\\
		62 & 67 & 0.38 & 77.34 & 18.68 & -4.52 & 3.55\\
		\addlinespace
		92 & 17 & 0.23 & 94.10 & 4.74 & -3.25 & 0.55\\
		71 & 45 & 0.33 & 69.95 & 12.55 & -2.48 & 2.07\\
		90 & 38 & 0.47 & 93.41 & 10.59 & -2.31 & 2.49\\
		84 & 43 & 0.46 & 89.26 & 11.99 & -2.30 & 2.76\\
		47 & 27 & 0.27 & 71.42 & 7.53 & -2.26 & 1.02\\
		\bottomrule
	\end{tabular}
	\hfill
	\begin{tabular}[t]{rrrrrrr}
		\toprule
		 & $d$ & $h$ & $c$ & $\hat{Q}_i$ & \multicolumn{1}{p{8ex}}{\raggedleft  Shapley value} & \multicolumn{1}{p{10ex}}{\raggedleft  hd-prop. value}\\
		\midrule
		40 & 36 & 0.43 & 84.91 & 10.04 & -1.71 & 2.16\\
		4 & 18 & 0.23 & 68.20 & 5.02 & -1.53 & 0.58\\
		78 & 98 & 0.37 & 61.09 & 27.32 & -1.45 & 5.05\\
		91 & 8 & 0.26 & 96.29 & 2.23 & -1.15 & 0.29\\
		76 & 20 & 0.43 & 86.34 & 5.58 & -0.54 & 1.20\\
		\addlinespace
		12 & 18 & 0.30 & 62.65 & 5.02 & -0.23 & 0.75\\
		16 & 95 & 0.34 & 52.38 & 26.49 & 0.42 & 4.50\\
		27 & 11 & 0.45 & 64.13 & 3.07 & 1.26 & 0.69\\
		34 & 85 & 0.49 & 69.76 & 23.70 & 1.91 & 5.81\\
		33 & 82 & 0.40 & 52.64 & 22.86 & 3.55 & 4.57\\
		\addlinespace
		85 & 163 & 0.46 & 58.40 & 45.44 & 5.46 & 10.45\\
		38 & 336 & 0.05 & 2.23 & 93.68 & 8.87 & 2.34\\
		23 & 481 & 0.09 & 7.93 & 134.10 & 11.90 & 6.03\\
		24 & 359 & 0.11 & 7.62 & 100.09 & 14.13 & 5.50\\
		50 & 313 & 0.10 & 1.49 & 87.26 & 18.97 & 4.36\\
		\addlinespace
		96 & 413 & 0.16 & 8.01 & 115.14 & 27.02 & 9.21\\
		94 & 387 & 0.12 & 1.08 & 107.89 & 28.01 & 6.47\\
		7 & 310 & 0.19 & 5.53 & 86.43 & 29.65 & 8.21\\
		36 & 363 & 0.19 & 6.84 & 101.20 & 32.13 & 9.61\\
		53 & 415 & 0.15 & 3.68 & 115.70 & 32.42 & 8.68\\
		\addlinespace
		44 & 397 & 0.16 & 3.02 & 110.68 & 34.47 & 8.85\\
		65 & 357 & 0.17 & 1.65 & 99.53 & 35.51 & 8.46\\
		41 & 494 & 0.15 & 4.57 & 137.73 & 36.04 & 10.33\\
		49 & 448 & 0.16 & 3.03 & 124.90 & 38.39 & 9.99\\
		37 & 377 & 0.23 & 7.83 & 105.11 & 40.00 & 12.09\\
		\addlinespace
		72 & 435 & 0.22 & 8.35 & 121.28 & 41.99 & 13.34\\
		100 & 420 & 0.23 & 6.67 & 117.10 & 46.05 & 13.47\\
		52 & 313 & 0.31 & 7.03 & 87.26 & 48.89 & 13.53\\
		89 & 310 & 0.33 & 8.06 & 86.43 & 50.48 & 14.26\\
		48 & 415 & 0.22 & 1.58 & 115.70 & 52.18 & 12.73\\
		\addlinespace
		66 & 351 & 0.33 & 8.92 & 97.86 & 54.98 & 16.15\\
		20 & 445 & 0.22 & 1.65 & 124.07 & 55.46 & 13.65\\
		25 & 332 & 0.36 & 9.75 & 92.56 & 56.62 & 16.66\\
		83 & 346 & 0.30 & 1.89 & 96.46 & 59.06 & 14.47\\
		74 & 320 & 0.44 & 9.84 & 89.22 & 68.19 & 19.63\\
		\addlinespace
		81 & 460 & 0.27 & 1.93 & 128.25 & 68.98 & 17.31\\
		9 & 303 & 0.48 & 8.84 & 84.48 & 72.68 & 20.27\\
		97 & 312 & 0.48 & 7.56 & 86.99 & 76.25 & 20.88\\
		46 & 334 & 0.43 & 4.56 & 93.12 & 76.54 & 20.02\\
		28 & 430 & 0.35 & 5.20 & 119.88 & 77.13 & 20.98\\
		\addlinespace
		70 & 467 & 0.35 & 7.79 & 130.20 & 78.03 & 22.78\\
		58 & 481 & 0.34 & 1.91 & 134.10 & 89.58 & 22.80\\
		45 & 429 & 0.38 & 1.52 & 119.60 & 90.50 & 22.72\\
		88 & 378 & 0.49 & 8.19 & 105.39 & 91.52 & 25.82\\
		21 & 352 & 0.49 & 2.48 & 98.14 & 94.40 & 24.04\\
		\addlinespace
		32 & 424 & 0.42 & 1.67 & 118.21 & 98.03 & 24.82\\
		1 & 419 & 0.45 & 4.03 & 116.82 & 99.27 & 26.28\\
		13 & 378 & 0.48 & 1.06 & 105.39 & 100.87 & 25.29\\
		6 & 430 & 0.47 & 1.18 & 119.88 & 111.23 & 28.17\\
		2 & 467 & 0.46 & 4.13 & 130.20 & 111.45 & 29.95\\
		\bottomrule
	\end{tabular}
	
	\caption{\label{t2} Parameters, optimal order sizes, Shapley value and hd-proportional value for Examples \ref{ex:2} and  \ref{ex:4} (items/firms sorted by Shapley values).}
\end{table}
\end{example}

\renewcommand{\baselinestretch}{1.5}

To conclude this section, we include a toy example that illustrates the interest that the use of the Shapley value can have for management decisions, specifically in the problem of choosing some items to stop distributing. Note that the Shapley value is the average of the marginal contribution of each item $i$ to coalitions $S\subseteq N\setminus\{i\}$. In particular, the Shapley value is also considering the marginal costs, i.e., the marginal contributions to the grand coalition $N$. For this reason, the Shapley value may be a better tool than marginal costs when deciding which set of items to stop distributing. 

\begin{example}
	\label{ex:2_1}
	We consider a firm that sells to its customers three types of items that it previously buys from a unique supplier. The firm distributes a total of nine items, three of each type, and has decided to stop distributing one item of each type. Let us consider that the corresponding inventory problem fits into a multi-item EOQ model with exemptable ordering costs. As in Example \ref{ex:2}, the ordering cost is EUR $2\,000$ and the firm is exempted from this charge for all orders greater than or equal to EUR $200\,000$. Table \ref{t1_1} shows for each of the items their corresponding parameters, i.e. their type, their monthly demands, their holding costs per unit and month, and their acquisition costs per unit. 
	
	\renewcommand{\baselinestretch}{1}
	
	\begin{table}
		\centering
		\fontsize{8}{10}\selectfont
		\begin{tabular}[t]{rrrrr}
			\toprule
			Type & Item & d & h & c\\
			\midrule
			1 & 1 & 37 & 0.48 & 58.61\\
			1 & 2 & 68 & 0.48 & 65.79\\
			1 & 3 & 57 & 0.46 & 90.21\\
			\bottomrule
		\end{tabular}
		\hfill
		\begin{tabular}[t]{rrrrr}
			\toprule
			Type & Item & d & h & c\\
			\midrule
			2 & 4 & 230 & 0.09 & 99.45\\
			2 & 5 & 245 & 0.05 & 66.12\\
			2 & 6 & 271 & 0.07 & 50.06\\
			\bottomrule
		\end{tabular}
		\hfill
		\begin{tabular}[t]{rrrrr}
			\toprule
			Type & Item & d & h & c\\
			\midrule
			3 & 7 & 423 & 0.29 & 9.93\\
			3 & 8 & 459 & 0.26 & 2.34\\
			3 & 9 & 429 & 0.29 & 1.44\\
			\bottomrule
		\end{tabular}
		
		\caption{\label{t1_1} Parameters for Example \ref{ex:2_1}.}
	\end{table}
	
	\renewcommand{\baselinestretch}{1.5}
	
	Taking into account that the firm will place joint orders for all items, the resulting optimal cycle length is $2.8443$ months and therefore the optimal number of orders placed per month is $0.3516$. Table \ref{t2_1} shows the parameters associated with all items, as well as their optimal order sizes. To decide which items of each type to stop distributing, the marginal costs and the Shapley value of each item are used to measure the importance of its contribution to the joint inventory cost. These data can be found in Table \ref{t2_1}. 	It can be seen that if marginal costs are used to make the decision, items $1$, $6$ and $9$ are no longer distributed, resulting in a joint inventory cost of EUR $618.61$. However, if the Shapley value is used, items $2$, $6$ and $9$ would be discontinued, resulting in a lower joint inventory cost: EUR $617.41$.

	\renewcommand{\baselinestretch}{1}
	
	\begin{table}
		\centering
		
		\fontsize{8}{10}\selectfont
		
		\begin{tabular}[t]{rrrrrrrr}
			\toprule
			Type & Item & d & h & c & $\hat{Q}_i$ & $c(N)-c(N\setminus\{i\})$ & \multicolumn{1}{p{8ex}}{\raggedleft  Shapley value}\\
			\midrule
			1 & 3 & 57 & 0.46 & 90.21 & 719.22 & -15.31 & 45.33\\
			1 & 1 & 37 & 0.48 & 58.61 & 700.25 & 3.66 & 48.99\\
			1 & 2 & 68 & 0.48 & 65.79 & 702.17 & 1.75 & 70.20\\
			\addlinespace
			2 & 4 & 230 & 0.09 & 99.45 & 999.66 & -295.75 & -214.19\\
			2 & 5 & 245 & 0.05 & 66.12 & 891.99 & -188.08 & -134.19\\
			2 & 6 & 271 & 0.07 & 50.06 & 838.76 & -134.85 & -82.46\\
			\addlinespace
			3 & 7 & 423 & 0.29 & 9.93 & 563.09 & 140.82 & 302.89\\
			3 & 8 & 459 & 0.26 & 2.34 & 542.48 & 161.43 & 325.61\\
			3 & 9 & 429 & 0.29 & 1.44 & 531.65 & 172.26 & 341.74\\
			\bottomrule
		\end{tabular}

		\caption{\label{t2_1} Parameters, optimal order sizes, marginal costs and Shapley value for Example \ref{ex:2_1} (items sorted by type and Shapley values).}
	\end{table}
	
	\renewcommand{\baselinestretch}{1.5}
\end{example}

\section{A problem with multiple firms}
\label{section4}
In this section we consider a situation where several firms place their orders jointly,  and introduce a model with several firms and one item for each firm: the {\em multi-firm EOQ problem with exemptable ordering costs}. The parameters that characterize such a problem are  the same as in the previous section, but their interpretation is different, which, as we will see, means that the approach from which we analyse the model in this section is essentially distinct. These parameters are as follows:

\begin{itemize}
	\item $N$ is the finite set of cooperating firms.
	\item $d_i>0$ is the demand that firm $i\in N$ has for its item per unit of time.
	\item $h_i>0$ is the cost for firm $i$ of holding one unit of its item for one unit of time.
	\item $c_i>0$ is the acquisition cost for firm $i$ of one unit of its item.
	\item $a>0$ is the ordering cost.
	\item $B > 0$ is the order price above which the supplier exempts the firms from the ordering cost.
\end{itemize}
We assume that the firms in $N$ has decided to place joint orders of their items. This means that, if we denote by $Q_i$ the order size of firm $i$, it holds that
\begin{equation}
\label{cycles_equality_2}
\frac{Q_i}{d_i}=\frac{Q_j}{d_j}, \mbox{ for all $i,j\in N$}.
\end{equation}
In such a case, by performing identical calculations to those in the previous section, we conclude that for a given coalition of cooperating firms $S\subseteq N$, the optimal order size for firm $i$ and the minimum average cost per time unit are those given in expressions (\ref{minimumCPT+}), (\ref{minimumvalueofCPT+}) and (\ref{minimumvalueofCPT+-}). Then, for every multi-firm EOQ problem with exemptable ordering costs given by $P=(N, \{d_i\}_{i\in N},\{h_i\}_{i\in N},\{c_i\}_{i\in N},a,B )$ we can again define its associated cost game $(N,c^P)$ as in (\ref{eq:game_item}), of which we know that it is strictly subadditive. In this context, we aim to distribute $CPT(\{\hat{Q}_j\}_{j\in N})$ in a way that is acceptable to all firms in $N$. 

Now, it is particularly relevant to identify some way of sharing the total cost that is coalitionally stable, i.e. that does not leave any subgroup of firms unsatisfied. In game theory terms, this is possible if the core of the cost game is non-empty. Below we give the definition of the core of $(N,c^P)$:
$$core(N,c^P)=\{x\in\mathbb{R}^N\ |\ \sum_{i\in N}x_i=c^P(N)\mbox{ and }\sum_{i\in S}x_i\leq c^P(S)\ \forall S\subseteq N\}.$$

The following example shows that the Shapley value of a cost game associated with a multi-firm EOQ problem with exemptable ordering costs may not belong to its core.

\begin{example}
	\label{ex:3}
	Take $P = (N,\{d_i\}_{i\in N},\{h_i\}_{i\in N},\{c_i\}_{i\in N},a,B)$ with $N=\{ 1,2,3\}$, $d=(1\,600,1\,700,1\,000)$, $h=(0.1,0.2,0.6)$, $c=(13,40,10)$, $a=6$, $B=3\,500$.
	Its associated cost game $(N,c^P)$ is given by
	\[\begin{array}{r|*{7}c}
	S & \{1\} & \{2\} & \{3\} & \{1,2\} & \{1,3\} & \{2,3\} & \{1,2,3\} \\ 
	\hline
	c^P(S) & 13.462 & 8.750 & 84.853 & 9.854 & 43.182 & 21.090 & 19.484 		
	\end{array}\]
	and $SH(N,c^P)=(-2.809,-16.211,38.504)$. Clearly $SH(N,c^P)\not\in core(N,c^P)$ because $SH_2(N,c^P)+SH_3(N,c^P)=22.293>21.090=c^P(\{2,3\}).$
\end{example}

We now introduce an allocation rule for multi-firm EOQ problems with exemptable ordering costs that always proposes core allocations of their associated cost games. To begin with, we give the formal definition of allocation rule in this context. We then define the the hd-proportional allocation rule. For this purpose, let us denote by $\mathcal{F}^N$  the family of multi-firm EOQ problems with exemptable ordering costs and set of firms $N$, and denote by $\mathcal{F}$ the family of all multi-firm EOQ problems with exemptable ordering costs.

\begin{definition}
	An allocation rule $\varphi$ for multi-firm EOQ problems with exemptable ordering costs is a map defined  on $\mathcal{F}$ that assigns to every $P=(N, \{d_i\}_{i\in N},\{h_i\}_{i\in N},\{c_i\}_{i\in N},a,B )\in \mathcal{F}$ a vector $\varphi(P)\in\mathbb{R}^N$ satisfying that $\sum_{i\in N}\varphi_i(P)=CPT(\{\hat{Q}_j\}_{j\in N})$.
\end{definition}

\begin{definition}
	The hd-proportional allocation rule $\varphi^{hd}$ for multi-firm EOQ problems with exemptable ordering costs is defined by
	\begin{equation*}
		\varphi^{hd}_i(P)=h_id_i  \min\left\{\dfrac{B}{2\sum_{j\in N}c_jd_j},\sqrt{\dfrac{2a}{\sum_{j\in N}h_jd_j}}\right\}.
	\end{equation*}
	for every $P=(N, \{d_i\}_{i\in N},\{h_i\}_{i\in N},\{c_i\}_{i\in N},a,B )\in \mathcal{F}$ and every $i\in N$.\footnote{This rule is similar to the SOC-rule introduced in \cite{Meca2004} for a different class of problems.}
\end{definition}

It is clear that $\varphi^{hd}$ is a well-defined allocation rule since $\sum_{i\in N}\varphi^{hd}_i(P)=CPT(\{\hat{Q}_j\}_{j\in N})$ for all $P\in\mathcal{F}$ (see, for instance (\ref{eq:game_item2})). Moreover, it always provides core allocations, as we prove below.

\begin{proposition}\label{thm:balan}
	Let $P = (N,\{d_i\}_{i\in N},\{h_i\}_{i\in N},\{c_i\}_{i\in N},a,B)$ be a multi-firm EOQ problem with exemptable ordering costs and let $(N, c^P)$ be its associated cost game. Then, $\varphi^{hd}(P)\in core(N, c^P)$.
\end{proposition}

\begin{proof}
Take a non-empty $ S\subseteq N$. Then
	\begin{eqnarray*}
		\sum_{i\in S}\varphi^{hd}_i(P)&=&  \sum_{i\in S} h_id_i \min\left\{\dfrac{B}{2\sum_{j\in N}c_jd_j},\sqrt{\dfrac{2a}{\sum_{j\in N}h_jd_j}}\right\}\\
		&<& \left(\sum_{j\in S}h_jd_j\right) \min\left\{\dfrac{B}{2\sum_{j\in S}c_jd_j},\sqrt{\dfrac{2a}{\sum_{j\in S}h_jd_j}}\right\}\\
		&=& c^P(S),
	\end{eqnarray*}
	where the inequality follows from $\sum_{j\in N}c_jd_j > \sum_{j\in S}c_jd_j$ and $\sum_{j\in N}h_jd_j > \sum_{j\in S}h_jd_j$.
\end{proof}

From what we have discussed so far it should be clear that the model studied in the previous section and the one studied in this section are the same from a mathematical point of view, but  different from the point of view of their interpretation and applicability in the context of real data. In the previous section, to compare the influence of the various items on the joint inventory cost we proposed to use a rule based on the Shapley value; in this section, to share the joint inventory cost when several firms cooperate, the hd-proportional allocation rule arises naturally. These are two different rules that we propose for different models; nevertheless, as mathematical objects, these two rules are comparable. And that is what we intend to do by returning to the example of the previous section.

\begin{example}
\label{ex:4}
Take the data from Table \ref{t1} in Example \ref{ex:2} in the previous section, but consider now that, instead of corresponding to one hundred items ordered by a single firm from the same supplier, they correspond to orders placed by one hundred different firms from the same supplier. Similarly to the Example \ref{ex:2}, the supplier is willing to meet any order placed jointly by the firms with an ordering cost of EUR $2\,000$, but all orders above or equal to EUR $200\,000$ would be exempt from this charge. This situation fits a multi-firm EOQ problem with exemptable ordering costs, and the optimal order sizes (now for every firm), the optimal cycle length and the optimal number of orders placed per month are the same as in Example \ref{ex:2}. However, it is now more appropriate to use a rule that results in core allocations, such as the hd-proportional rule, to apportion joint inventory costs. Table \ref{t2} above shows in its last column the proposed distribution according to the hd-proportional rule. Note that the proposals provided by the Shapley  and the hd-proportional rules are quite different. This is not surprising considering that the Shapley value is designed to produce fair shares, while the hd-proportional rule is designed to produce coalitionally stable shares. Figure \ref{f2} shows such differences graphically. In the figure, the line of circles shows the Shapley value of the items and the line of triangles shows the proposal of the hd-proportional rule for the firms. The items or firms, according to the interpretation of each example, are renumbered in increasing order of their Shapley values. A simple glance allows to see that the hd-proportional rule has a lower variability than the Shapley value; moreover, the rankings generated by both rules are different.
\begin{figure}
	
	{\centering \includegraphics[width=0.8\linewidth]{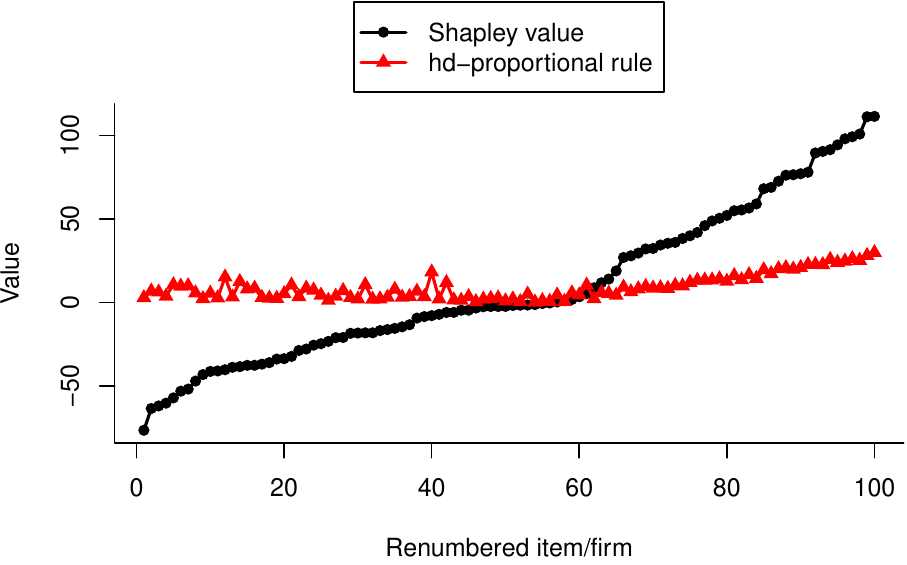} }
	
	\caption{Shapley value vs hd-proportional rule: items/firms renumbered in increasing order of their Shapley values.}\label{f2}
\end{figure}

\end{example}

In view of the results so far in this section, we recommend the use of the hd-proportional rule for the allocation of optimal inventory costs in a multi-firm EOQ problem with exemptable ordering costs, mainly because it provides coalitionally stable allocations. In the remainder of this section we will further study its properties and provide an axiomatic characterisation for it. 

The first property is stability, which requires that an allocation rule  proposes allocations that do not leave any subset of the set of firms unsatisfied, i.e. that it  proposes core allocations.

\bigskip
\noindent
\textbf{Stability:} An allocation rule $\varphi$ satisfies stability if  for every $P\in \mathcal{F}^N$ and every $S\subseteq N$, then 
$$\sum_{i\in S}\varphi_i(P)\leq CPT(\{\hat{Q}_j\}_{j\in S}).$$

Bearing in mind that $CPT(\{\hat{Q}_j\}_{j\in S})=c^P(S)$, Proposition \ref{thm:balan} implies that $\varphi^{hd}$ satisfies stability. The next property we consider simply indicates that an allocation rule must allocate a non-negative amount for all firms, since they all contribute a non-negative quantity to the joint inventory cost.

\bigskip
\noindent
\textbf{Non-negativity:} An allocation rule $\varphi$ satisfies non-negativity if  for every $P\in \mathcal{F}^N$ and every $i\in N$, then $\varphi_i(P)\geq 0$.

\bigskip
It is clear that $\varphi^{hd}$ is a non-negative rule. We now introduce two properties that we will use to characterise the hd-proportional rule. Both are related to the magnitudes $h_id_i$ which, for each firm $i$, give the holding cost per unit of time of the demand per unit of time of firm $i$, which, in short, we will call the holding cost of the demand of $i$. The first property indicates that if two firms have the same holding cost of the demand, then the rule should assign them the same amount.

\bigskip
\noindent
\textbf{Symmetry in the holding costs of the demands:} An allocation rule $\varphi$ satisfies symmetry in the holding costs of the demands if,  for every $P\in \mathcal{F}^N$ and every $i,j\in N$ with $h_id_i=h_jd_j$, then $\varphi_i(P)=\varphi_j(P)$.

\bigskip
Finally, the non-manipulability property requires that if several firms merge into one firm whose holding costs of the demands and whose acquisition costs are respectively the sum of the costs of the demands and the sum of the acquisition costs of the merging firms, then the rule must allocate to the merged firm the sum of what it allocated before the merger to the merging firms.

\bigskip
\noindent
\textbf{Non-manipulability:} An allocation rule $\varphi$ satisfies non-manipulability if, for every pair $P=(N, \{d_i\}_{i\in N},\{h_i\}_{i\in N},\{c_i\}_{i\in N},a,B )$ and
$ P^{\prime}=(N^{\prime}, \{d_i^{\prime}\}_{i\in N^{\prime}},\{h^\prime_i\}_{i\in N^{\prime}},\{c^{\prime}_i\}_{i\in N^{\prime}},a,B )$ in $\mathcal{F}$ with  $N^{\prime}\subseteq N$, and with $i\in N^{\prime}$ such that $h^{\prime}_id^{\prime}_i=h_id_i+\sum_{j\in N\setminus N^{\prime}}h_jd_j$ and $c^{\prime}_id^{\prime}_i=c_id_i+\sum_{j\in N\setminus N^{\prime}}c_jd_j$,   and satisfying that, for every $j\in N^{\prime}\setminus\{i\}$, $h^{\prime}_jd^{\prime}_j=h_jd_j$ and $c^{\prime}_jd^{\prime}_j=c_jd_j$,  then
\begin{equation*}
\varphi_i(P^{\prime})=	\varphi_i(P)+\sum_{j\in N\setminus N^{\prime}}	\varphi_j(P).
\end{equation*}

As mentioned above, the hd-proportional rule is a stable allocation rule. Moreover, the other three properties we have introduced characterise it univocally, as stated in Theorem \ref{theo:hd} below. The proof of this result is quite laborious and, moreover, it is an adaptation to this context of the proof of the main result in \cite{Moreno-Ternero2006}. We have therefore chosen to include it in a separate appendix.

\begin{theorem}\label{theo:hd}
	The hd-proportional rule is the unique allocation rule for multi-firm EOQ problems with exemptable ordering costs  that satisfies non-negativity, symmetry in the holding costs of the demands and non-manipulability.
\end{theorem}

\section{A problem with multiple firms and items}
A generalisation that arises naturally from the models in the previous sections is the {\em multi-firm-item EOQ problem with exemptable ordering costs}, in which several firms  each distributing several items cooperate to order all their items together. The parameters that characterise such a problem are as follows:

\begin{itemize}
	\item $M$ is the finite set of cooperating firms.
	\item For every firm $k\in M$, $N_k$ is the finite set of the items it orders. $N$ denotes the set of all ordered items $\cup_{k\in M}N_k$.
	\item $d_{ki}>0$ is the demand per unit of time of item $i\in N_k$, for every firm $k\in M$.
	\item $h_{ki}>0$ is the cost of holding one unit of item $i\in N_k$ for one unit of time, for every firm $k\in M$.
	\item $c_{ki}>0$ is the acquisition cost of one unit of item $i\in N_k$, for every firm $k\in M$.
	\item $a>0$ is the ordering cost.
	\item $B > 0$ is the order price above which the supplier exempts the firms from the ordering cost.
\end{itemize}

Although this model is more general than those presented in the previous sections, it is  nothing more than a multi-item EOQ problem with exemptable ordering costs in which there is also a partition of the set of items, where each class contains the items of one of the firms involved. Seen in this light, all the results of Section \ref{sectionthree} still hold true in this context. In particular, when the firms in $M$ cooperate and place joint orders for all their items, then 
the quotients $Q_{ki}/d_{ki}$  should be equal for every $k\in M$ and  $i\in N_k$, where $Q_{ki}$ denotes the order size for item $i$ of firm $k$. Moreover, the optimal order size for $k$ and $i$ is given by
\begin{equation}
\label{minimumCPT++}
\hat{Q}_{ki}=\left\{
\begin{array}{ll}
\sqrt{\frac{2ad_{ki}^2}{\sum_{l\in M}\sum_{j\in N_l}h_{lj}d_{lj}}}&
\mbox{if $2\sqrt{\frac{2a}{\sum_{l\in M}\sum_{j\in N_l}h_{lj}d_{lj}}}<\frac{B}{\sum_{l\in M}\sum_{j\in N_l}c_{lj}d_{lj}},$}\\\\
\frac{B d_{ki}}{\sum_{l\in M}\sum_{j\in N_l}c_{lj}d_{lj}}&
\mbox{if $2\sqrt{\frac{2a}{\sum_{l\in M}\sum_{j\in N_l}h_{lj}d_{lj}}}\geq\frac{B}{\sum_{l\in M}\sum_{j\in N_l}c_{lj}d_{lj}}.$}\\
\end{array}
\right.
\end{equation}
Furthermore the minimum average cost per time unit is given by
\begin{equation}
\label{minimumvalueofCPT++}
CPT(\hat{Q}_{ki})=\left\{
\begin{array}{ll}
\sqrt{2a\sum_{l\in M}\sum_{j\in N_l}h_{lj}d_{lj}}&\mbox{if $2\sqrt{\frac{2a}{\sum_{l\in M}\sum_{j\in N_l}h_{lj}d_{lj}}}<\frac{B}{\sum_{l\in M}\sum_{j\in N_l}c_{lj}d_{lj}},$}\\\\
\frac{B}{2}\frac{\sum_{l\in M}\sum_{j\in N_l}h_{lj}d_{lj}}{\sum_{l\in M}\sum_{j\in N_l}c_{lj}d_{lj}}&\mbox{if $2\sqrt{\frac{2a}{\sum_{l\in M}\sum_{j\in N_l}h_{lj}d_{lj}}}\geq\frac{B}{\sum_{l\in M}\sum_{j\in N_l}c_{lj}d_{lj}},$}\\
\end{array}
\right.
\end{equation}
and it is clear that $CPT(\hat{Q}_{ki})$ can also be written as
\begin{equation}
\label{minimumvalueofCPT+++}
CPT(\hat{Q}_{ki})=\sum_{l\in M}\sum_{j\in N_l}h_{lj}d_{lj}  \min\left\{\dfrac{B}{2\sum_{l\in M}\sum_{j\in N_l}c_{lj}d_{lj}},\sqrt{\dfrac{2a}{\sum_{l\in M}\sum_{j\in N_l}h_{lj}d_{lj}}}\right\}.
\end{equation}
Notice that $CPT(\hat{Q}_{ki})$ is the total average cost per unit of time incurred by all items and firms and does not depend on particular $k$ and $i$. Therefore, we denote this quantity interchangeably by $CPT(\hat{Q}_{ki})$ or by $CPT(\{\hat{Q}_{lj}\}_{l\in M, j\in N_l})$. On some occasions it will also be in our interest to fix $S\subseteq M$ and calculate $CPT(\{\hat{Q}_{lj}\}_{l\in S, j\in N_l})$ by changing $S$ to $M$ in (\ref{minimumvalueofCPT++}) or (\ref{minimumvalueofCPT+++}).

Following the guidelines of the previous sections, we now propose to design an allocation rule for multi-firm-item EOQ problems with exemptable ordering costs, which allocates the total cost resulting from the cooperation between the firms in a coalitionally stable way and includes a procedure to evaluate the impact of each item on this total cost.

Formally, a multi-firm-item EOQ problem with exemptable ordering costs $\Pi$ is given by the seven components listed at the beginning of this section:
\begin{equation}
\label{mfieoq}
\Pi=(M, \{N_k\}_{k\in M}, \{d_{ki}\}_{k\in M, i\in N_k},\{h_{ki}\}_{k\in M, i\in N_k},\{c_{ki}\}_{k\in M, i\in N_k},a,B ).
\end{equation}
We denote by 
$\mathcal{FI}^{M,N}$  the family of multi-firm-item EOQ problems with exemptable ordering costs and sets of firms and items $M$ and $N$; we denote by $\mathcal{FI}$ the family of all multi-firm-item EOQ problems with exemptable ordering costs.

\begin{definition}
	An allocation rule $\varPhi$ for multi-firm-item EOQ problems with exemptable ordering costs is a map defined  on $\mathcal{FI}$ that assigns to every $\Pi\in \mathcal{FI}^{M,N}$ a vector $\varPhi(\Pi)\in\mathbb{R}^N$ satisfying that $\sum_{i\in N}\varPhi_i(\Pi)=CPT(\{\hat{Q}_{lj}\}_{l\in M, j\in N_l})$.
\end{definition}

A first property we want for an allocation rule in this context is {\em stability for firms}, which requires it to propose allocations that do not leave any subset of firms unsatisfied.

\bigskip
\noindent
\textbf{Stability for firms:} An allocation rule $\varPhi$ satisfies stability for firms if for every $\Pi\in \mathcal{FI}^{M,N}$ and every $S\subseteq M$, then 
$$\sum_{l\in S}\sum_{j\in N_l}\varPhi_j(\Pi)\leq CPT(\{\hat{Q}_{lj}\}_{l\in S, j\in N_l}).$$

Our objective is to construct an allocation rule that satisfies stability for firms and that allows each firm to assess the impact of each of its items on the cost that it has to pay according to the rule. To achieve this, we define the rule in two phases. In a first phase we distribute the total cost among the firms; in a second phase we distribute the amount allocated to each firm in the first phase among its items. 

Let us formally define our rule, which we call the Shapley-proportional allocation rule and denote by $\varPhi^{sp}$. In the first phase, to allocate the total cost across firms we rely on the hd-proportional rule from the previous section. Take $\Pi\in \mathcal{FI}$ given by (\ref{mfieoq}) and, for every firm $k\in M$, denote by $\varPhi^{sp}_{[k]}(\Pi)$ the sum $\sum_{i\in N_k}\varPhi^{sp}_{i}(\Pi)$. We then ask $\varPhi^{sp}$ to comply
\begin{equation}
\label{quotient1}
\varPhi^{sp}_{[k]}(\Pi)=\sum_{i\in N_k}h_{ki}d_{ki}  \min\left\{\dfrac{B}{2\sum_{l\in M}\sum_{j\in N_l}c_{lj}d_{lj}},\sqrt{\dfrac{2a}{\sum_{l\in M}\sum_{j\in N_l}h_{lj}d_{lj}}}\right\}.
\end{equation}
In view of Proposition \ref{thm:balan} it is easy to check that a rule  defined using (\ref{quotient1}) satisfies  stability for firms. 
Now, in a second phase, we have to complete the definition of the rule for allocating each $\varPhi^{sp}_{[k]}(\Pi)$ among the items of firm $k$. Following the principles used in Section \ref{sectionthree}, we use the Shapley value to make such an allocation. For every firm $k\in M$ and every coalition $S\subseteq N_k$ define the  multi-firm-item EOQ problem with exemptable ordering $\Pi^{k,S}$ given by
\begin{equation*}
\label{m-f-i}
\Pi^{k,S}=(M, \{\bar{N}_l\}_{l\in M}, \{d_{lj}\}_{l\in M, j\in \bar{N}_l},\{h_{lj}\}_{l\in M, j\in \bar{N}_l},\{c_{lj}\}_{l\in M, j\in \bar{N}_l},a,B ),
\end{equation*}
where $\bar{N}_k=S$ and $\bar{N}_l=N_l$ for all $l\in M\setminus\{k \}$. Now define the cost game $(N_k,c^{\Pi,k})$ by
$$c^{\Pi,k}(S)=\varPhi^{sp}_{[k]}(\Pi^{k,S})$$
for all $S\subseteq N_k$. Finally, the Shapley-proportional allocation rule is defined for every $i\in N_k$ and every $k\in M$ by
\begin{equation}
\label{rulefinal}
\varPhi^{sp}_i(\Pi)=SH_i(N_k,c^{\Pi,k}).
\end{equation}
Notice that $\varPhi^{sp}$ defined by (\ref{rulefinal}) satisfies (\ref{quotient1}) and, then, stability for firms. Besides, in view of (\ref{minimumvalueofCPT+++}) it is clear that $\varPhi^{sp}$ is well defined, i.e., that $\sum_{i\in N}\varPhi^{sp}_i(\Pi)=CPT(\{\hat{Q}_{lj}\}_{l\in M, j\in N_l}$. Furthermore, the two following statements hold true for each multi-firm-item EOQ problem with exemptable ordering $\Pi\in \mathcal{FI}^{M,N}$ given by (\ref{mfieoq}):
\begin{enumerate}
	\item If $|M|=1$ and $M=\{ k\}$, then it holds that $\varPhi^{sp}(\Pi)=SH(N,c^P)$, where $P$ is the multi-item problem defined by $( N_k, \{d_{ki}\}_{i\in N_k},\{h_{ki}\}_{i\in N_k},\{c_{ki}\}_{i\in N_k},a,B ).$
	\item If $|N_k|=1$ for all $k\in M$, then it holds that $\varPhi^{sp}(\Pi)=\varphi^{hd}(P)$, where $P$ is the multi-firm problem defined by $( \cup_{k\in M}N_k, \{d_{ki}\}_{k\in M,i\in N_k},\{h_{ki}\}_{k\in M,i\in N_k},\{c_{ki}\}_{k\in M,i\in N_k},a,B ).$
\end{enumerate}

This two-step procedure for constructing the Shapley-proportional allocation rule is similar to that used in \cite{Owen1977} for constructing a value for cooperative games with unions. In Owen's words, his purpose was to modify the Shapley value of a cooperative game ``so as to take into account the possibility that some players --because of personal or political affinities-- may be more likely to act together than others''. Formally, Owen starts from a cooperative game and a partition of the set of players whose classes, ``unions'', contain affine players. In this context, Owen modifies the Shapley value to distribute among the players what they are all able to generate, taking into account the aforementioned affinities. Owen introduces his value using a two-step procedure and characterizes it by appropriately modifying the properties that characterize the Shapley value. This approach of Owen's has been widely used later in the literature to propose alternative modifications of the Shapley value for games with unions (see, for example \citealp{Kamijo2009}) and to propose modifications of other values for games with unions (e.g., of the Banzhaf value in \citealp{Alonso2002}, of the $\tau$-value in \citealp{Casas2003}). However, our approach is novel because we do not intend to adapt a pre-existing value to the case with unions, but rather to apportion the total cost among the unions in a stable way and to assess the contribution of each player to the total cost taking into account the mode of apportionment among unions that has been adopted.

In the following we will give a characterisation of the Shapley-proportional rule. The properties we will use are adaptations of those characterising the hd-proportional rule for multi-firm problems and, in addition, a new property related to the Shapley value. This characterization is far from those of values for games with unions, because it combines two different approaches: that of stability for the unions and that of fairness for the players within each union. 

\bigskip
\noindent
\textbf{Non-negativity for firms:} An allocation rule $\varPhi$ satisfies non-negativity for firms if, for every $\Pi\in \mathcal{FI}^{M,N}$ and every $k\in M$, then $\sum_{i\in N_k}\varPhi_{i}(\Pi)\geq 0$.

\bigskip
\noindent
\textbf{Symmetry in the holding costs of the demands for firms:} An allocation rule $\varPhi$ satisfies symmetry in the holding costs of the demands for firms if, for every $\Pi\in \mathcal{FI}^{M,N}$ and every $k,l\in M$ with $\sum_{i\in N_k}h_{ki}d_{ki}=\sum_{j\in N_l}h_{lj}d_{lj}$, then $\sum_{i\in N_k}\varPhi_i(\Pi)=\sum_{j\in N_l}\varPhi_j(\Pi)$.

\bigskip
\noindent
\textbf{Non-manipulability for firms:} An allocation rule $\varPhi$ satisfies non-manipulability for firms if, for every pair of problems in $\mathcal{FI}$
$$\Pi=(M, \{N_k\}_{k\in M}, \{d_{ki}\}_{k\in M, i\in N_k},\{h_{ki}\}_{k\in M, i\in N_k},\{c_{ki}\}_{k\in M, i\in N_k},a,B )$$
$$\Pi^{\prime}=(M^{\prime}, \{N^{\prime}_k\}_{k\in M^{\prime}}, \{d^{\prime}_{ki}\}_{k\in M^{\prime}, i\in N^{\prime}_k},\{h^{\prime}_{ki}\}_{k\in M^{\prime}, i\in N^{\prime}_k},\{c^{\prime}_{ki}\}_{k\in M^{\prime}, i\in N^{\prime}_k},a,B )$$ 
with  $M^{\prime}\subseteq M$, and with $k\in M^{\prime}$ such that $\sum_{i\in N^{\prime}_k}h^{\prime}_{ki}d^{\prime}_{ki}=\sum_{i\in N_k}h_{ki}d_{ki}+\sum_{l\in M\setminus M^{\prime}}\sum_{j\in N_l}h_{lj}d_{lj}$ and  $\sum_{i\in N^{\prime}_k}c^{\prime}_{ki}d^{\prime}_{ki}=\sum_{i\in N_k}c_{ki}d_{ki}+\sum_{l\in M\setminus M^{\prime}}\sum_{j\in N_l}c_{lj}d_{lj}$,   and satisfying that, for every $l\in M^{\prime}\setminus\{k\}$, $\sum_{j\in N^{\prime}_l}h^{\prime}_{lj}d^{\prime}_{lj}=\sum_{j\in N_l}h_{lj}d_{lj}$ and $\sum_{j\in N^{\prime}_l}c^{\prime}_{lj}d^{\prime}_{lj}=\sum_{j\in N_l}c_{lj}d_{lj}$,  then
\begin{equation*}
	\sum_{i\in N^{\prime}_k}\varPhi_i(\Pi^{\prime})=	\sum_{i\in N_k}\varPhi_i(\Pi)+\sum_{l\in M\setminus M^{\prime}}\sum_{j\in N_l}	\varPhi_j(\Pi).
\end{equation*}

As we have already indicated, these three properties are adaptations of those in Section \ref{section4}. In the following, we present an adaptation to this context of the balanced contributions property which, as is well known, is fulfilled by the Shapley value (see, for example, \citealp{Myerson1980}). This property is desirable to ensure the fairness of the assessment of the impact of the items on the cost to be paid by the firm. It means that the effect on the assessment of item $i$ if item $j$ is removed from the orders is the same as the effect on the assessment of item $j$ if item $i$ is removed from the orders (for any items $i$ and $j$ from the same firm).

\bigskip
\noindent
\textbf{Balanced contributions for items within each firm:} An allocation rule $\varPhi$ satisfies balanced contributions for items within each firm if for every $\Pi\in \mathcal{FI}^{M,N}$, every $k\in M$ and every $i,j\in N_k$, then $$\varPhi_{i}(\Pi)-\varPhi_{i}(\Pi_{-j})=\varPhi_{j}(\Pi)-\varPhi_{j}(\Pi_{-i}),$$
where, for any item $r$ of firm $l$,  $\Pi_{-r}$ denotes an identical problem  to $\Pi$ except that $r$ has been removed from $N_{l}$.

\begin{theorem}\label{theo:shp}
The Shapley-proportional rule is the unique allocation rule for multi-firm-item EOQ problems with exemptable ordering costs  that satisfies non-negativity for firms, symmetry in the holding costs of the demands for firms, non-manipulability for firms and balanced contributions for items within each firm.
\end{theorem}
\begin{proof} It is clear that the Shapley-proportional rule $\varPhi^{sp}$ satisfies the properties. To prove the uniqueness take a rule $\varPhi$ that also satisfies them. Since $\varPhi$ and $\varPhi^{sp}$ satisfy  non-negativity for firms, symmetry in the holding costs of the demands for firms, and non-manipulability for firms, then it is easy to check that the uniqueness in Theorem \ref{theo:hd} implies
\begin{equation}
\label{conteq}
\sum_{i\in N_k}\varPhi_i(\Pi)=\sum_{i\in N_k}\varPhi^{sp}_i(\Pi)
\end{equation}
for every $\Pi\in \mathcal{FI}^{M,N}$ and every $k\in M$. To conclude the proof take  $\Pi\in \mathcal{FI}^{M,N}$ and $k\in M$; we prove that $\varPhi_i(\Pi)=\varPhi^{sp}_i(\Pi)$ for every $i\in N_k$ by induction in $|N_k|$. If $|N_k|=1$, then (\ref{conteq}) implies that $\varPhi_i(\Pi)=\varPhi^{sp}_i(\Pi)$ for the unique $i\in N_k$. Assume that the equality is true for every $i\in N_k$ when $|N_k|\leq l$. Consider now the case $|N_k|=l+1$ and take $i,j\in N_k$. Then
\begin{equation}
\label{conteq2}
\varPhi_{i}(\Pi)-\varPhi_{j}(\Pi)=\varPhi_{i}(\Pi_{-j})-\varPhi_{j}(\Pi_{-i})=\varPhi_{i}^{sp}(\Pi_{-j})-\varPhi_{j}^{sp}(\Pi_{-i})=\varPhi_{i}^{sp}(\Pi)-\varPhi_{j}^{sp}(\Pi),
\end{equation}
where the first and third equalities hold because $\varPhi$ and $\varPhi^{sp}$ respectively satisfy balanced contributions for items within each firm, and the second because of the induction hypothesis. 
Thus, 
\begin{equation}
	\label{conteq3}
	\varPhi_{i}(\Pi)-\varPhi_{i}^{sp}(\Pi)=\varPhi_{j}(\Pi)-\varPhi_{j}^{sp}(\Pi).
\end{equation}
Since (\ref{conteq3}) holds true for every $i,j\in N_k$, then (\ref{conteq}) implies that 
$$\varPhi_{i}(\Pi)=\varPhi_{i}^{sp}(\Pi)$$
for all $i\in N_k$, and the proof is concluded.
\end{proof}

To finish this section we return to the data from Example \ref{ex:2} to illustrate the behaviour of the Shapley-proportional rule.

\begin{example}
\label{ex:5}
We take again the data from Example \ref{ex:2} and consider now that they correspond to one hundred items belonging to eight firms that place joint orders. Table \ref{tab:data_sec5} shows the items, indicating in brackets the firm to which they belong, and the allocations for each of them of the Shapley value, the hd-proportional rule and the Shapley-proportional rule. The items are arranged in increasing order of the allocation assigned to them by the Shapley-proportional rule within each firm. We observe that the three give rise to different rankings of the items and that the allocations of the hd-proportional and Shapley-proportional rules are closer to each other, while the allocation given by the Shapley value is considerably farther apart. In this example, what each firm would have to pay according to the Shapley-proportional rule is the sum of the allocations of this rule to the firm's items, as shown in Table \ref{tab:firm_ex2}.

\renewcommand{\baselinestretch}{1}

\begin{table}
	\fontsize{8}{10}\selectfont
	\begin{tabular}[t]{lrrr}
		\toprule
		\multicolumn{1}{p{8ex}}{\raggedright Item\\ (Firm)} & \multicolumn{1}{p{8ex}}{\raggedleft Shapley\\ } & \multicolumn{1}{p{9ex}}{\raggedleft hd-prop. rule} & \multicolumn{1}{p{12ex}}{\raggedleft Shapley-prop.\\ rule}\\
		\midrule
		47(1) & -2.26 & 1.02 & 0.83\\
		67(1) & -40.86 & 2.85 & 1.32\\
		90(1) & -2.31 & 2.49 & 2.19\\
		93(1) & -8.34 & 3.56 & 2.99\\
		24(1) & 14.13 & 5.50 & 5.52\\
		\addlinespace
		23(1) & 11.90 & 6.03 & 5.94\\
		14(1) & -9.26 & 6.78 & 5.97\\
		31(1) & -32.18 & 10.49 & 8.77\\
		36(1) & 32.13 & 9.61 & 9.95\\
		68(1) & -7.79 & 18.28 & 16.89\\
		\addlinespace
		74(1) & 68.19 & 19.63 & 20.57\\
		46(1) & 76.54 & 20.02 & 21.21\\
		70(1) & 78.03 & 22.78 & 23.88\\
		58(1) & 89.58 & 22.80 & 24.26\\
		21(1) & 94.40 & 24.04 & 25.60\\
		\addlinespace
		92(2) & -3.25 & 0.55 & 0.44\\
		76(2) & -0.54 & 1.20 & 1.12\\
		43(2) & -76.38 & 2.98 & 1.30\\
		59(2) & -16.74 & 2.24 & 1.79\\
		15(2) & -18.37 & 3.01 & 2.51\\
		\addlinespace
		94(2) & 28.01 & 6.47 & 6.78\\
		7(2) & 29.65 & 8.21 & 8.50\\
		30(2) & -51.73 & 10.07 & 8.70\\
		48(2) & 52.18 & 12.73 & 13.35\\
		66(2) & 54.98 & 16.15 & 16.74\\
		\addlinespace
		28(2) & 77.13 & 20.98 & 21.90\\
		6(2) & 111.23 & 28.17 & 29.62\\
		26(3) & -4.54 & 1.38 & 1.21\\
		19(3) & -18.07 & 1.99 & 1.51\\
		5(3) & -38.71 & 3.33 & 2.38\\
		\addlinespace
		55(3) & -28.63 & 3.55 & 2.82\\
		16(3) & 0.42 & 4.50 & 4.37\\
		33(3) & 3.55 & 4.57 & 4.50\\
		50(3) & 18.97 & 4.36 & 4.64\\
		3(3) & -37.55 & 8.16 & 7.15\\
		\addlinespace
		37(3) & 40.00 & 12.09 & 12.69\\
		69(3) & -40.13 & 15.28 & 14.10\\
		100(3) & 46.05 & 13.47 & 14.18\\
		20(3) & 55.46 & 13.65 & 14.57\\
		83(3) & 59.06 & 14.47 & 15.46\\
		\addlinespace
		9(3) & 72.68 & 20.27 & 21.49\\
		22(4) & -43.08 & 2.36 & 2.10\\
		40(4) & -1.71 & 2.16 & 2.23\\
		63(4) & -36.79 & 3.03 & 2.86\\
		77(4) & -16.11 & 3.18 & 3.18\\
		\addlinespace
		11(4) & -60.14 & 3.80 & 3.47\\
		51(4) & -24.57 & 4.46 & 4.45\\
		34(4) & 1.91 & 5.81 & 6.09\\
		95(4) & -56.99 & 10.67 & 10.71\\
		80(4) & -18.08 & 10.66 & 11.02\\
		\bottomrule
	\end{tabular}
	\hfill
	\begin{tabular}[t]{lrrr}
		\toprule
		\multicolumn{1}{p{8ex}}{\raggedright Item\\ (Firm)} & \multicolumn{1}{p{8ex}}{\raggedleft Shapley\\} & \multicolumn{1}{p{9ex}}{\raggedleft hd-prop. rule} & \multicolumn{1}{p{12ex}}{\raggedleft Shapley-prop.\\ rule}\\
		\midrule
		39(5) & -35.91 & 2.66 & 1.75\\
		60(5) & -7.01 & 2.22 & 1.95\\
		61(5) & -14.48 & 3.23 & 2.75\\
		18(5) & -33.52 & 5.32 & 4.36\\
		54(5) & -15.40 & 7.95 & 7.28\\
		\addlinespace
		65(5) & 35.51 & 8.46 & 8.91\\
		44(5) & 34.47 & 8.85 & 9.27\\
		96(5) & 27.02 & 9.21 & 9.45\\
		10(5) & -5.87 & 12.02 & 11.44\\
		72(5) & 41.99 & 13.34 & 13.79\\
		\addlinespace
		52(5) & 48.89 & 13.53 & 14.13\\
		25(5) & 56.62 & 16.66 & 17.35\\
		97(5) & 76.25 & 20.88 & 21.90\\
		91(6) & -1.15 & 0.29 & 0.24\\
		4(6) & -1.53 & 0.58 & 0.51\\
		\addlinespace
		12(6) & -0.23 & 0.75 & 0.71\\
		8(6) & -23.18 & 1.36 & 0.84\\
		35(6) & -5.83 & 1.64 & 1.45\\
		82(6) & -18.21 & 2.13 & 1.69\\
		71(6) & -2.48 & 2.07 & 1.94\\
		\addlinespace
		86(6) & -46.94 & 5.78 & 4.69\\
		57(6) & -37.50 & 8.61 & 7.67\\
		85(6) & 5.46 & 10.45 & 10.35\\
		98(6) & -38.30 & 12.58 & 11.58\\
		81(6) & 68.98 & 17.31 & 18.45\\
		\addlinespace
		32(6) & 98.03 & 24.82 & 26.50\\
		13(6) & 100.87 & 25.29 & 27.03\\
		38(7) & 8.87 & 2.34 & 2.42\\
		73(7) & -20.97 & 3.80 & 2.82\\
		99(7) & -13.21 & 3.63 & 2.90\\
		\addlinespace
		87(7) & -41.17 & 5.64 & 3.91\\
		78(7) & -1.45 & 5.05 & 4.62\\
		64(7) & -20.87 & 7.16 & 5.98\\
		29(7) & -25.45 & 7.47 & 6.14\\
		75(7) & -52.98 & 9.81 & 7.47\\
		\addlinespace
		53(7) & 32.42 & 8.68 & 9.14\\
		49(7) & 38.39 & 9.99 & 10.58\\
		41(7) & 36.04 & 10.33 & 10.82\\
		45(7) & 90.50 & 22.72 & 24.40\\
		88(7) & 91.52 & 25.82 & 27.39\\
		\addlinespace
		1(7) & 99.27 & 26.28 & 28.09\\
		2(7) & 111.45 & 29.95 & 32.00\\
		27(8) & 1.26 & 0.69 & 0.71\\
		42(8) & -33.78 & 2.52 & 2.31\\
		84(8) & -2.30 & 2.76 & 2.80\\
		\addlinespace
		62(8) & -4.52 & 3.55 & 3.59\\
		79(8) & -61.83 & 6.30 & 5.96\\
		17(8) & -63.35 & 6.89 & 6.54\\
		56(8) & -27.76 & 8.63 & 8.63\\
		89(8) & 50.48 & 14.26 & 15.06\\
		\bottomrule
	\end{tabular}
	\caption{\label{tab:data_sec5} Shapley value, hd-proportional rule and Shapley-proportional rule for Examples \ref{ex:2}, \ref{ex:4} and \ref{ex:5} (ordered by Shapley-proportional rule and firm).}
\end{table}

\renewcommand{\baselinestretch}{1.5}

\begin{table}
	\centering
	\fontsize{10}{12}\selectfont
	\begin{tabular}[t]{c|rrrrrrrr}
		\toprule
		Firm & 1&2&3&4&5&6&7&8\\
		\midrule
		Shapley-prop. rule & 175.89& 112.75& 121.07& 46.13& 124.34& 113.67& 178.68& 45.59\\
		\bottomrule
	\end{tabular}
	
	\caption{\label{tab:firm_ex2} The allocations of the Shapley-proportional rule to firms in Example \ref{ex:5}.}
\end{table}

\end{example}

\section{Concluding remarks and further research}

In this paper we introduced and analysed several continuous review inventory models in which the firms are exempted from ordering costs if the price of their orders is greater than or equal to a certain amount. These inventory models are based on the activity of an electrical material distribution firm that contacted one of the authors for advice. 

The problems that the firm addresses are of a new type that we called EOQ problems with exemptable ordering costs. We first considered a simple model with one firm and one item. We then formulated a basic EOQ problem with exemptable ordering costs for which we obtained the optimal ordering policy.  

Second, we studied a model with one firm and several items for which we obtained an optimal ordering policy and a procedure based on the Shapley value to evaluate the impact of each item on the total cost.  We called it a multi-item EOQ problem with exemptable ordering costs. We quantitatively illustrated this multi-item model by showing the case study of a firm that sells to its customers one hundred items that it previously buys from a unique supplier. We obtained the optimal order plan for this firm taking into account that it places joint orders for all items. To help the manager make certain decisions such as, for example, to stop distributing some items, we proposed  the Shapley value of each item as a measure of the importance of each item's contribution to the joint inventory cost. In this case study, we observed that the items that contribute most to the joint inventory cost tend to have high demands and holding costs, but low adquisition costs.  

Third, we considered a situation where several firms place their orders jointly  and we called it  a multi-firm problem with exemptable ordering costs. We obtained an optimal ordering policy for this problem and a procedure, the hd-proportional value, which allocates the total cost among the firms proportionally to the holding costs of the demands. This procedure allocates  the total cost in a coalitionally stable way. We also provided an axiomatic characterization of our procedure. Both models, multi-item and multi-firm,  are the same from a mathematical point of view, but  different from the point of view of their interpretation and applicability in the context of real data.  In the multi-item model, to compare the influence of the various items on the joint inventory cost we proposed to use the Shapley value. However, in the multi-firm model, to allocate the joint inventory cost when several firms cooperate, the hd-proportional value arose as a matter of course. To compare these two values, we decided to re-interpret the previous case study, so that instead of one firm buying one hundred items from the supplier, we considered one hundred firms each buying a single item from the same supplier. In this new scenario, we showed that the Shapley value and the hd-proportional value propositions were quite different. In this example, we observe that the hd-proportional value has lower variability than the Shapley value and, moreover, the rankings generated by the two values are different. This is not surprising considering that the Shapley value was designed to provide fair allocations, while the hd-proportional value was designed to provide coalitionally stable allocations.  

Fourth, we considered a new model, which is a natural generalisation of the two previous ones, in which several firms  each distributing several items cooperate to order all their items together. We called it a
multi-firm-item EOQ problem with exemptable ordering costs.  We obtained an optimal ordering policy for this model and a rule, the Shapley-proportional rule, to allocate the total cost among the firms in a coalitionally stable way and to assess the impact of each item on the cost that its firm would have to pay. We also provided an axiomatic characterisation of this rule and calculated it for the case study data, now considering that the one hundred items belong to eight different firms. From the comparison of the allocations of the Shapley value, the hd-proportional rule and the Shapley-proportional rule, we observed that all three result in different rankings of the items, which is to be expected. 

To conclude, we would like to comment that our models have some limitations that can be addressed in future research. We can look first a multi-item EOQ problem with exemptable ordering costs and with upper bounded order sizes. In this model with capacity constrains, apart from deriving the optimal order sizes and  proposing a sharing rule, it would be interesting to identify an optimal ordering partition of the set of items. Second, researchers could analyse a multi-firm EOQ problem with exemptable ordering costs and with fuzzy demand. This model would suit certain situations in which firms may not have a regular and predictable demand, but they may have an estimate of a certain range in which demand occurs. Knowing the optimal order sizes and allocating the total inventory cost among the firms in an uncertain environment needs further study. Third, it would also be interesting to study a multi-firm EOQ problem with exemptable ordering costs and with acquisition costs per unit depending on the order sizes. If the acquisition cost decreases as the order size increases, we would be in a multi-firm EOQ problem with exemptable ordering costs and other discounts. In this model with reduced costs, apart from deriving the optimal order sizes, it would be interesting to find a coalitionally stable allocation rule that would incentivise firms to place larger orders in order to obtain large discounts that will reduce their  costs. Finally, we could extend our model by considering that each item/firm $i$ has a different ordering cost $a_i$. Depending on how we aggregate the individual ordering costs, i.e., whether we define the ordering cost of a coalition $S \subseteq N$ by $a_S= \sum_{i \in S} a_i$ (as do \citealp{Anily2007}, \citealp{Dror2007}, and \citealp{Dror2012}), by $a_S=max_{i \in S}{a_i}$ (as do \citealp{Fiestras2012}), or by a general function of $(a_i)_{i \in S}$ (as does \citealp{Saavedra2020}), would result in different models that require further study. However, these three approaches have in common that they will generally entail the loss of the subadditivity property of the corresponding games, which will raise challenging research questions.

\section*{Appendix}

The following is a proof of Theorem \ref{theo:hd}. For this we need to introduce some notation and some properties, as well as to prove some previous lemmas.

With respect to the notation, take a multi-firm EOQ problem with exemptable ordering costs $P=(N, \{d_i\}_{i\in N},\{h_i\}_{i\in N},\{c_i\}_{i\in N},a,B )$. For every $i\in N$, denote by $H_i$ the product $h_id_i$ and by $C_i$ the product $c_id_i$. 

Let us now introduce two new properties for allocation rules in this context.

\bigskip
\noindent
\textbf{Strong non-manipulability:} An allocation rule $\varphi$ satisfies strong non-manipulability if  for every pair $P=(N, \{d_i\}_{i\in N},\{h_i\}_{i\in N},\{c_i\}_{i\in N},a,B )$ and
$ P^{\prime}=(N^{\prime}, \{d_i^{\prime}\}_{i\in N^{\prime}},\{h^\prime_i\}_{i\in N^{\prime}},\{c^{\prime}_i\}_{i\in N^{\prime}},a,B )$ in  $\mathcal{F}$ satisfying that  $N^{\prime}\subseteq N$ and that there exists $i\in N^{\prime}$ such that $\HDip=\HDi+\sum_{j\in N\setminus N^{\prime}}\HDj$, that $\CDip=\CDi+\sum_{j\in N\setminus N^{\prime}}\CDj$,  and that, for every $j\in N^{\prime}\setminus\{i\}$, $\HDjp=\HDj$ and $\CDjp=\CDj$,  then it holds that
$\varphi_j(P^{\prime})=\varphi_j(P), \text{ for every }  j\in N^{\prime}\setminus\{i\}$.

\bigskip
\noindent
\textbf{hd-ranking preservation:} An allocation rule $\varphi$ satisfies hd-ranking preservation if  for every $P=(N, \{d_i\}_{i\in N},\{h_i\}_{i\in N},\{c_i\}_{i\in N},a,B )\in \mathcal{F}$ and every $i,j\in N$ with $\HDi>\HDj$, then $\varphi_i(P)\geq\varphi_j(P)$.

\begin{lemma}\label{lem:relp1}
	Take $\varphi$ an allocation rule  for multi-firm EOQ problems with exemptable ordering costs. Then, $\varphi$ satisfies strong  non-manipulability if and only if $\varphi$ satisfies non-manipulability.
\end{lemma}
\begin{proof}
It is clear that if $\varphi$ satisfies strong  non-manipulability then it also satisfies non-manipulability. Conversely,  let us assume that $\varphi$ satisfies the property of non-manipulability and take $P, P^{\prime}, N^{\prime}$ and $i\in N^{\prime}$ as in the statement of such a property. For any 
	 $j\in N^{\prime}\setminus\{i\}$ define 
	 $$P^{\prime\prime}=(N^{\prime}\setminus \{j\}, \{d_k^{\prime\prime}\}_{k\in N^{\prime}\setminus \{j\}},\{h_k^{\prime\prime}\}_{k\in N^{\prime}\setminus \{j\}},\{c_k^{\prime\prime}\}_{k\in N^{\prime}\setminus \{j\}},a,B )\in \mathcal{F}$$ with $\HDipp=\HDip+\HDjp$, $\CDipp=\CDip+\CDjp$, and, for every $k\in N^{\prime}\setminus\{i,j\}$, $\HDkpp=\HDkp$, $\CDkpp=\CDkp$. Applying the property of non-manipulability of $\varphi$ to the problems $P$, $P^{\prime}$ and $P^{\prime\prime}$, we have
	\begin{equation}\label{eq:pppp}
		\varphi_i(P^{\prime\prime})=\varphi_i(P^{\prime})+\varphi_j(P^{\prime})=	\varphi_i(P)+\sum_{k\in N\setminus N^{\prime}}	\varphi_k(P)+\varphi_j(P^{\prime}).
	\end{equation}
	Now, applying the property of non-manipulability of $\varphi$ to  situations $P$ and $P^{\prime\prime}$, we have
	\begin{equation}\label{eq:ppp}
		\varphi_i(P^{\prime\prime})=\varphi_i(P)+\varphi_j(P)+\sum_{k\in N\setminus N^{\prime}}\varphi_k(P).
	\end{equation}
	Then, comparing Equation~(\ref{eq:pppp}) and Equation~(\ref{eq:ppp}), we obtain $\varphi_j(P)=\varphi_j(P^{\prime})$ and the proof is concluded.
\end{proof}

\begin{lemma}\label{lem:relp3}
Take $\varphi$ an allocation rule  for multi-firm EOQ problems with exemptable ordering costs. Then, if $\varphi$ satisfies non-negativity, non-manipulability and symmetry in the holding costs of the demands, then it also satisfies hd-ranking preservation.
\end{lemma}
\begin{proof}
	Take $P=(N, \{d_l\}_{l\in N},\{h_l\}_{l\in N},\{c_l\}_{l\in N},a,B )\in \mathcal{F}$ and $i,j\in N$ with $\HDi>\HDj$. Take $k\not\in N$ and define
	$$P^\prime=(N^{\prime}, \{d_l^{\prime}\}_{l\in N^{\prime}},\{h^\prime_l\}_{l\in N^{\prime}},\{c^{\prime}_l\}_{l\in N^{\prime}},a,B )\in \mathcal{F}$$ 
	where $N^{\prime}=N\cup\{k\}$, and it is satisfied that $\HDlp=\HDl$ and $\CDlp=\CDl$ for every $l\in N\setminus\{i\}$, 
	$\HDip=\HDj$, $\HDkp=\HDi-\HDj$, and $\CDip+\CDkp=\CDi$ with $\CDip>0$, and $\CDkp>0$. Clearly, $c^P(N)=c^{P^{\prime}}(N\cup\{k\})$. Since $\varphi$ satisfies non-manipulability, we have
	\begin{equation*}
		\varphi_i(P)=\varphi_i(P^{\prime})+\varphi_k(P^{\prime}).
	\end{equation*}
	Besides, by symmetry in the holding costs of the demands, $\varphi_i(P^{\prime})=\varphi_j(P^{\prime})$. Then, using that $\varphi$ satisfies also strong non-manipulability, it holds $\varphi_j(P^{\prime})=\varphi_j(P)$. Thus,
	\begin{equation*}
		\varphi_i(P)=\varphi_j(P)+\varphi_k(P^{\prime})\geq \varphi_j(P),
	\end{equation*}
	where the inequality follows from the non-negative property of  $\varphi$.
\end{proof}

\noindent
{\bf Proof of Theorem \ref{theo:hd}.}
	It is clear that the hd-proportional rule satisfies non-negativity,  symmetry in the holding costs of the demands and non-manipulability.
	Conversely, take a  rule $\varphi$ that satisfies non-negativity, symmetry in the holding costs of the demands and non-manipulability. From Lemma~\ref{lem:relp1}, and Lemma~\ref{lem:relp3}, it also satisfies strong non-manipulability and hd-ranking preservation. Let  $P=(N, \{d_i\}_{i\in N},\{h_i\}_{i\in N},\{c_i\}_{i\in N},a,B )\in \mathcal{F}$. We show that $\varphi(P)=\HD(P)$. We consider two situations:
	\begin{itemize}
	\item [Case 1.] $|N|=2$. For simplicity take $N=\{i,j\}$. If $\HDi=\HDj$, by  symmetry in the holding costs of the demands, we have $\HD_i(P)=\varphi_i(P)=\varphi_j(P)=\HD_j(P)$. Let us assume  $\HDi>\HDj$ and take $k\in \mathbb{N}$  such that $\HDi/k<\HDj$ and $\HDj=[\frac{\HDj k}{\HDi}]\frac{\HDi}{k}+r_j$ with $0<r_j<\frac{\HDi}{k}$ (where $[ \frac{\HDj k}{\HDi}]$ denotes the integer part of $ \frac{\HDj k}{\HDi}$). Define the problem $P^{\prime}$ where $a$ and $B$ have the same values as in $P$ and
		\begin{itemize}
			\item $N^{\prime}=\{j,i_1,\ldots,i_k\}$
			\item $\HDjp=\HDj$, $\HDilp=\HDi/k$,  for every $l=1,\ldots,k$.
			\item $\CDjp=\CDj$, $\CDi=\sum_{l=1}^k\CDilp$ with $\CDilp>0$ for every $l=1,\ldots,k$.
		\end{itemize}
		Then, by non-manipulability, $\varphi_i(P)=\sum_{l=1}^k\varphi_{i_l}(P^{\prime})$ and by symmetry in the holding costs of the demands, we have $\varphi_{i_l}(P^{\prime})=\varphi_i(P)/k$ and  $\varphi_j(P)=\varphi_j(P^{\prime})$.
		
		Take  $q=[\frac{\HDj k}{\HDi}]\geq 1$ and $P^{\prime\prime}\in \mathcal{F}$ defined by
		\begin{itemize}
			\item  $N^{\prime\prime}=\{i_1,\ldots,i_k\}\cup \{j_1,\ldots, j_q,j_{q+1}\}$,
			\item $\HDilpp=\frac{\HDi}{k}=\HDjhpp$, for every $l=1,\ldots,k$, $h=1,\ldots,q$,
			\item $\HDjqpp=r_j$, 
			\item $\CDi=\sum_{l=1}^k\CDilp$, and  $\CDj=\sum_{h=1}^{q+1}\CDjhpp$, with $\CDjhpp>0$ for every $h=1,\ldots,q+1$,
		\end{itemize}
		and with $a$ and $B$ equal to those of $P$. The symmetry in the holding costs of the demands implies that 
	\begin{equation}
		\label{11111}\varphi_{i_l}(P^{\prime\prime})=\varphi_{j_h}(P^{\prime\prime})
	\end{equation}
		for every $l=1,\ldots,k$ and every $h=1,\ldots,q$. The strong non-manipulability implies that
		\begin{equation}
			\label{22222}
			\varphi_{i_l}(P^{\prime\prime})=\varphi_{i_l}(P^{\prime})=\frac{\varphi_{i}(P)}{k}
		\end{equation}
		for every $l=1,\ldots,k$. Now by non-manipulability,  non-negativity and the relationship between the values of the rule $\varphi$ in the situations $P$ and $P^{\prime}$, we have
		\begin{equation}\label{eq:des1}
			\varphi_j(P)=\varphi_j(P^{\prime})=\sum_{h=1}^{q+1}\varphi_{j_{h}}(P^{\prime\prime})=\frac{q}{k}\varphi_i(P)+\varphi_{j_{q+1}}(P^{\prime\prime})\geq \frac{q}{k}\varphi_i(P).
		\end{equation}
		Now take into account that $\HDjqpp=r_j$ and that $r_j<\frac{\HDi}{k}=H_{j_h}^{\prime\prime}$ for any $h\in\{ 1,\ldots ,q\}$. Then, the
	property hd-ranking preservation implies that $\varphi_{j_{q+1}}(P^{\prime\prime})\leq \varphi_{j_{h}}(P^{\prime\prime})$ for any $h\in\{ 1,\ldots ,q\}$. Now (\ref{11111}), (\ref{22222})  and (\ref{eq:des1}) implies that
		\begin{equation}\label{eq:des2}
			\varphi_j(P)=\frac{q}{k}\varphi_i(P)+\varphi_{j_{q+1}}(P^{\prime\prime})\leq \frac{q+1}{k}\varphi_i(P).
		\end{equation}
		By the choice of $q$, we have 
		\begin{equation*}
			\frac{\HDj k}{\HDi}-1\leq q\leq \frac{\HDj k}{\HDi}\quad\text{or equivalently}\quad 	\frac{\HDj}{\HDi}-\frac{1}{k}\leq \frac{q}{k}\leq \frac{\HDj}{\HDi}.
		\end{equation*}
		Thus, using the non-negativity property of the rule and inequalities~(\ref{eq:des1})  and (\ref{eq:des2}), we have
		\begin{equation*}
			\left(\frac{\HDj}{\HDi}-\frac{1}{k}\right)\varphi_i(P)\leq \varphi_j(P)\leq \left(\frac{\HDj}{\HDi}+\frac{1}{k}\right)\varphi_i(P).
		\end{equation*}
		Since this is true for every $k^{\prime}\in\mathbb{N}$ with $k^{\prime}>k$, then $	\frac{\HDj}{\HDi}\varphi_i(P)=\varphi_j(P)$ and $\varphi(P)=\HD(P)$.
		
		\item [Case 2.] $|N|>2$. Take $i\in N$ and $j\in N\setminus{i}$. We define a new situation $P^{\prime}$ as follows:
		\begin{itemize}
			\item $N^{\prime}=\{i,j\}$,
			\item  $\HDip=\HDi$, $\HDjp=\sum_{k\in N\setminus\{i\}}\HDj$,
			\item $\CDip=\CDi$, $\CDjp=\sum_{k\in N\setminus\{i\}}\CDj$,
		\end{itemize}
		where $a$ and $B$ have the same values as in $P$. By Case 1 we know that
		\begin{equation*}
			\HDi\varphi_j(P^{\prime})=\varphi_i(P^{\prime})\sum_{k\in N\setminus\{i\}}\HDj.
		\end{equation*}
		By strong non-manipulability, we have $\varphi_i(P^{\prime})=\varphi_i(P)$ and   $\varphi_j(P^{\prime})=c^P(N)-\varphi_i(P)$. Thus,
		\begin{equation*}
			\HDi(c^P(N)-\varphi_i(P))=\varphi_i(P)\sum_{k\in N\setminus\{i\}}\HDj
		\end{equation*}
		and
		\begin{equation*}
			\varphi_i(P)=\frac{\HDi}{\sum_{j\in N}\HDj}c^P(N)=\HD_i(P).
		\end{equation*}
		Since $i$ is an arbitrary firm, we obtain $\varphi_j(P)=\HD_j(P)$, for every $j\in N$. $\square$
	\end{itemize}

\end{document}